\documentclass[11pt,english,%
titlepage,%
]{report}
\usepackage[letterpaper,left=1in,right=1in,top=1in,bottom=1in]{geometry}
\usepackage[utf8]{inputenc}

\usepackage{graphicx}
\usepackage{multirow}
\usepackage{import}
\usepackage{standalone}
\usepackage{siunitx}
\usepackage{fancyhdr}
\usepackage{pdfpages}
\usepackage{amsmath,amssymb,amsthm}
\usepackage{verbatim}
\usepackage{titlesec}
\usepackage{enumitem}
\usepackage{wrapfig}
\usepackage{subcaption}

\usepackage{algorithm}
\usepackage{algpseudocode}
\usepackage{amsmath}

\usepackage{anyfontsize}
\usepackage{tikz}
\usetikzlibrary{shapes.geometric,shapes.symbols,arrows.meta,positioning,fit,backgrounds,decorations.pathreplacing,calc}

\definecolor{sunfill}{RGB}{255,236,204}
\definecolor{sunedge}{RGB}{201,124,20}
\definecolor{roi}{RGB}{31,102,168}
\definecolor{ext}{RGB}{176,42,88}
\definecolor{impl}{RGB}{118,118,126}
\definecolor{ink}{RGB}{40,44,52}

\definecolor{taskblue}{RGB}{50,100,200}
\definecolor{suborange}{RGB}{245,150,50}
\definecolor{eggshell}{RGB}{255,249,230}
\definecolor{lightblue}{RGB}{190,220,255}
\definecolor{lightaqua}{RGB}{205,245,245}

\usepackage{pgfplots}
\pgfplotsset{compat=1.16}

\usepackage[blocks]{authblk} 

\usepackage[style=ieee, maxbibnames=99]{biblatex}
\usepackage{url}
\usepackage{hyperref}
\hypersetup{
    colorlinks=true,
    linkcolor=blue,
    filecolor=magenta,      
    urlcolor=cyan,
}

\usepackage{xcolor}
\usepackage{graphicx}
\usepackage{booktabs} 

\newcommand{\chapfnt}{\fontsize{12}{17}}
\newcommand{\secfnt}{\fontsize{12}{17}}

\usepackage{etoolbox}
\makeatletter
\patchcmd{\chapter}{\if@openright\cleardoublepage\else\clearpage\fi}{}{}{}
\makeatother

\titleformat{\chapter}
{\normalfont\chapfnt\bfseries}{\thechapter}{1em}{\chapfnt}

\titleformat{\section}
{\normalfont\secfnt\bfseries}{\thesection}{1em}{\secfnt}

\titleformat{\subsection}
{\normalfont\secfnt\bfseries}{\thesubsection}{1em}{\secfnt}

\titlespacing\chapter{0pt}{2pt plus 1pt minus 1pt}{1pt plus 2pt minus 2pt}
\titlespacing\section{0pt}{2pt plus 1pt minus 1pt}{1pt plus 1pt minus 1pt}
\titlespacing\subsection{0pt}{2pt plus 1pt minus 1pt}{1pt plus 1pt minus 1pt}
\titlespacing\subsubsection{0pt}{2pt plus 1pt minus 1pt}{1pt plus 1pt minus 1pt}

\title{SWAN-SF-V2: A Multivariate Time Series Dataset for Space Weather Data Analytics; Now with AIA and Weighted SHARP Parameters}

\author[1]{Dustin Kempton}
\author[2]{Griffin Goodwin}
\author[1]{Reet Gupta}
\author[2]{Viacheslav M Sadykov}
\author[1]{Rafal A. Angryk}
\author[1]{Berkay Aydin}
\author[3]{Manolis K. Georgoulis}

\affil[1]{Department of Computer Science, Georgia State University \thanks{dkempton1@gsu.edu}}
\affil[2]{Department of Physics \& Astronomy, Georgia State University}
\affil[3]{Applied Physics Laboratory, Johns Hopkins University}

\date{\today}

\begin{document}

\maketitle
\thispagestyle{fancy}

\begin{abstract}
Since its original publication in 2020, the SWAN-SF dataset has been cited in over 300 research publications. Here we describe a comprehensive update to the original multivariate time series (MVTS) dataset extracted from solar photospheric vector magnetograms in the Spaceweather HMI Active Region Patch (SHARP) series. As was performed in the original release, our dataset includes a cross-checked NOAA solar flare catalog that facilitates the comparative evaluation of methods designed for solar flare prediction. We discuss the methods used for data collection, cleaning, and pre-processing of the solar active region and flare data, which includes improvements to the data integration and sampling methodology. Additional parameters have been derived from the Solar Dynamics Observatory's Atmospheric Imaging Assembly to augment the original features derived from magnetogram data.  Our dataset expands the period of observations to cover 7,730 MVTS data collections from active regions occurring between May 2010 and December 2025, each of which is composed of 182 flare-predictive parameters. We continue to integrate flare reports, with over 39,000 multi-source verified flare reports providing flaring history information about the source active regions. The primary tasks enabled by the disseminated dataset continue to include optimization of solar flare prediction and detailed investigation for elusive flare predictors or precursors, with both operational (research-to-operations), and basic research (operations-to-research) benefits.

\end{abstract}


\pagenumbering{roman}
\setcounter{page}{1}
\tableofcontents
\fancypagestyle{plain}{%
\fancyfoot[C]{Table of Contents}
}
\newpage

\pagenumbering{arabic}
\setcounter{page}{1}
\fancyfoot[C]{SWAN-SF-V2}
\fancyfoot[R]{Page \thepage}

\fancypagestyle{plain}{%
\fancyfoot[C]{SWAN-SF-V2}
\fancyfoot[R]{Page \thepage}
}

\chapter{Background \& Summary}

Space weather research has been recognized by several administrations as a topic of strategic national importance, from a 2008 report by the National Research Council concluding that a solar superstorm similar to the 1859 Carrington event \cite{carrington1859description} could disrupt the entire US power grid for months, and cause economic damage of $\sim$\$1-2\,T \cite{nrc2009}, to ``The White House'' releasing the National Space Weather Strategy and Space Weather Action Plan in 2015\cite{nswap2015}, with additional plans released in 2019 \cite{Hutson2019} and 2022 \cite{swr202rframework2022sworm}. More recently, the National Science and Technology Council explicitly called to ``Support development of forecasting capabilities based on statistical analysis, machine learning, and other innovative data science techniques'' in 2023 \cite{nswap2023}. As mentioned in \cite{enpSWE, nita:2018:roadmap}, a key component for this approach is to produce benchmark datasets for testing flare prediction algorithms.

As was the case in the original SWAN-SF dataset release \cite{angryk2020multivariate, angryk2020multivariatedata}, the benchmark dataset described in this work (hereafter, SWAN-SF-V2) is intended as a standardized collection for solar physicists and/or machine learning practitioners to compare and track the performance of machine learning models. Its purpose is to provide a machine learning ready dataset that is cleaned, integrated, and readily available, with data verified from multiple sources. The primary focus of the dataset is, again, solar flares and, by proxy, coronal mass ejections (CMEs) \cite{Benz2008, howard2011coronal, martens2017data, Ji2025}. These events occur in the solar corona and heliosphere, and can have a major negative impact on our technology-dependent society \cite{nswap2015}. A flare, being characterized by a sudden brightening, by orders of magnitude, in Extreme Ultra-Violet (EUV), X-ray, and, for large events, gamma-ray emissions, from a small area on the Sun, is a key driver in our augmentation of the features released in the original dataset with observations from the Solar Dynamics Observatory's Atmospheric Imaging Assembly \cite{lemen:sdoAIA}.  

Successful flare predictions via machine learning models trained and tested on this dataset intend to (1) tackle a central problem in space weather forecasting and (2) help identify physical mechanisms pertaining, or even giving rise, to solar flares. Like the original release, this dataset is intended to be a reliable resource for providing an unbiased comparison between results from various flare prediction algorithms. Producing a dataset of this nature is crucial for advancements in this domain, because without the use of a fixed dataset, such as the one presented here, discrepancies in performance evaluation metrics between different machine learning methods cannot be attributed unambiguously to the differences in the dataset or the quality of the methods at hand.

Since the magnetic field in active regions (ARs) is the only reservoir capable of storing sufficient energy to support solar transient events (STEs) such as solar flares, the magnetic field data are widely used for building STE forecasts \cite{ahmed2013solar,bobra2015solar,bobra2016, falconer2011tool}. The Helioseismic Magnetic Imager (HMI) \cite{schou2012design} onboard the Solar Dynamics Observatory (SDO) \cite{pesnell2015solar} observes the Sun almost continuously, and provides information on the magnetic field in the solar photosphere. However, since flares are known to originate from active regions (ARs; namely, areas of high concentration of magnetic flux), not the entire solar disk but its sub-portions are relevant. Thus, HMI active region patches were first created {\cite{Hoeksema2014}}. The HARP is a data pipeline product that identifies and tracks ARs in the solar photosphere, providing trimmed vector magnetic field maps. HARPs were then enriched with metadata (i.e., physical parameters inferred by magnetograms) of space weather forecasting interest, giving rise to Spaceweather HARPs, or SHARPs {\cite{bobra2014helioseismic}}. Therefore, the benchmark dataset mainly relies on Spaceweather HMI Active Region Patches (SHARPs) \cite{bobra2014helioseismic}, available from the Joint Science Operations Center (JSOC). Information on possible flares occurring in the region of interest, however, is missing from the SHARPs. 

For information about flare occurrences, we turn to the National Oceanic and Atmospheric Administration (NOAA), which operates Geostationary Operational Environmental Satellites (GOES) that, among others, have X-ray and particle detectors onboard. Since 1975, GOES have been detecting solar flares, and a catalog of all detected flares is available from NOAA \cite{swpcdata} while flare reports are also available through other resources, such as the Heliophysics Events Knowledgebase (HEK) \cite{hurlburt:hek}. These flares are classified logarithmically via their peak X-ray flux as A, B, C, M and X. The GOES flare catalog contains the flare time (start, peak, end), GOES class, peak X-ray flux, a spatial location on the solar disk (where available), and associated NOAA AR number (where available). Additionally, the Solar Region Summary (SRS) product Prepared jointly by NOAA's Space Weather Prediction Center and the U.S. Air Force, provides daily data on numbered ARs, including mean location and sunspot classification.

Additional flare reports have also been automatically detected by various solar feature detection modules \cite{Milligan2018}, and are regularly collected in various databases. These modules include Flare Detective {\cite{Martens2012}}, SSW Latest Events {\cite{SSWonline}}, RHESSI {\cite{Su2006}} and Hinode flare observations {\cite{Watanabe2012}}, and some others. Reports from two of these modules, SSW Latest Events and Hinode Flare Catalog, are used as auxiliary data sources to verify the missing locations of flares in the GOES catalog. Combining NOAA AR location information and auxiliary flare data sources enables the recovery and verification of missing locations for flares in the GOES catalog. The methods utilized in the process of cleaning, verifying, and combining the individual source data are described in the following section.

\chapter{Methods}

The methods utilized in creating this benchmark dataset for solar flare prediction follow closely the methods described in our original release \cite{angryk2020multivariate, angryk2020multivariatedata}. The dataset is still primarily based on photospheric magnetograms of the Sun's surface, with the addition of cutouts from the AIA instrument that are derived from the HARP magnetic map location information. So, we now approach the dataset construction as a four-fold problem. First, solar flare reports from GOES need to be cleaned, with conflicting information resolved. Second, solar flare event records need to be matched with regions of interest on the solar disk, i.e., active regions. Third, cutouts are made to intersect with the HARP solar magnetic data and produce image texture data of the various AIA wavelengths. Finally, sampling biases need to be eliminated when creating labeled datasets for training machine learning models. 

A deviation from the original methodology occurs in the way solar flare reports are matched with regions of interest on the solar disk, so we begin by describing our flare report integration in Section~\ref{sec:flareinteg} below. We then follow that with describing the SHARP data and flare integration into the resultant multi-variate dataset in Section~\ref{sec:cleanmvts}. This is then followed by a description of the AIA parameters that are the largest addition to the new version of the dataset in Section~\ref{section:aia_params}. 

\begin{figure}[htb!]
    \centering
    \begin{tikzpicture}[
      font=\sffamily,
      db/.style={
        cylinder, shape border rotate=90, aspect=0.28,
        draw=black!60, thick,
        cylinder uses custom fill, cylinder body fill=black!4, cylinder end fill=black!9,
        align=center, font=\fontsize{8}{8}\sffamily, text width=2.3cm, inner sep=2pt},
      stepA/.style={
        rectangle, rounded corners=3pt, draw=red!45!black, thick,
        fill=red!22, align=center, font=\sffamily,
        text width=2.3cm, minimum height=1.6cm, inner sep=4pt},
      stepB/.style={
        rectangle, rounded corners=3pt, draw=purple!50!black, thick,
        fill=purple!12, align=center, font=\sffamily,
        text width=2.3cm, minimum height=1.6cm, inner sep=4pt},
      stepC/.style={
        rectangle, rounded corners=3pt, draw=blue!55!black, thick,
        fill=blue!28, align=center, font=\sffamily, text=blue!25!black,
        text width=2.2cm, minimum height=1.6cm, inner sep=4pt},
      doc/.style={
        tape, tape bend top=none, tape bend bottom=in and out,
        draw=teal!60!black, thick, fill=teal!14,
        align=center, font=\sffamily\small,
        minimum width=2.5cm, minimum height=1.7cm, inner sep=3pt},
      green db/.style={
        cylinder, shape border rotate=90, aspect=0.26,
        draw=green!40!black, very thick,
        cylinder uses custom fill, cylinder body fill=green!16, cylinder end fill=green!26,
        align=center, font=\sffamily\bfseries, text=green!25!black,
        text width=2.2cm, inner sep=5pt},
      orange db/.style={
        cylinder, shape border rotate=90, aspect=0.30,
        draw=orange!65!black, thick,
        cylinder uses custom fill, cylinder body fill=orange!55, cylinder end fill=orange!38,
        text width=1.05cm, inner sep=2pt},
      groupA/.style={
        rectangle, rounded corners=14pt, draw=red!50!black, thick, dashed,
        fill=red!12, inner sep=7pt},
      groupB/.style={
        rectangle, rounded corners=14pt, draw=orange!60!black, thick, fill=orange!18, inner sep=3pt},
      arr/.style={-{Stealth[length=3mm,width=2.6mm]}, line width=0.9pt, draw=black},
    ]
    
    \node[db] (goesflr) at (1.1,11.45) {NOAA/GOES\\Flare Reports};
    \node[db] (ssw)     at (1.1, 9.95) {SSW Latest\\Events\\Flare Reports};
    \node[db] (xrt)     at (1.1, 8.40) {Hinode/XRT\\Flare Reports};
    \node[db] (srs)     at (1.1, 6.85) {NOAA/SRS\\Daily Active\\Region Reports};
    \node[db] (sharps)  at (1.1, 5.35) {SDO/HMI\\SHARPs};
    \node[db] (aiad)  at (1.1, 3.75) {SDO/AIA\\L1.5 Data};
    
    \node[stepA] (arc) at (4.25,11.45) {Active Region\\Centroid\\Augmentation};
    \node[stepA] (srsc) at (4.25,7.4) {SRS Outlier\\Correction};
    
    \node[stepB] (fm)  at ( 7.45,11.45) {Flare Matching\\on Time and\\Magnitude};
    \node[stepB] (dv)  at (10.65,11.45) {Distance-based\\Verification};
    \node[stepB] (sla) at (13.80,11.45) {Secondary\\Location\\Augmentation};
    \node[font=\sffamily, text=red!35!black] (xtitle) at (10.65,12.70)
          {Flare Cross-checking from Secondary Sources};
    
    \begin{scope}[on background layer]
      \node[groupA, fit=(fm)(dv)(sla)(xtitle)] (xgroup) {};
    \end{scope}
    
    \node[doc] (docc) at (12.46,8.76) {};
    \node[doc] (docb) at (12.28,8.58) {};
    \node[doc] (doca) at (12.10,8.40) {Cleaned and\\cross-checked\\flare reports};
    
    \node[stepC] (mfp) at (4.25,5.35) {Magnetic Field\\Parameter\\Generation};
    \node[stepC] (aia) at (7.25,5.35) {AIA Texture Parameter\\Integration};
    \node[stepC] (fhp) at (10.25,5.35) {Flare History\\Parameter\\Integration};
    
    \node[green db] (mvts) at (15.0,2.10) {MVTS Dataset\\for Space\\Weather\\Analytics};
    
    \node[orange db] (o1) at ( 5.45,1.45) {\rule{0pt}{9mm}};
    \node[orange db] (o2) at ( 6.80,1.45) {\rule{0pt}{9mm}};
    \node[orange db] (o3) at ( 8.15,1.45) {\rule{0pt}{9mm}};
    \node[orange db] (o4) at ( 9.50,1.45) {\rule{0pt}{9mm}};
    \node[orange db] (o5) at (10.85,1.45) {\rule{0pt}{9mm}};
    \node[align=center, font=\sffamily, text=orange!40!black] (mltitle) at (8.15,3.00)
          {Machine-learning-ready Dataset Creation\\
           Partitioning, Slicing \& Labeling, Undersampling};
    
    \begin{scope}[on background layer]
      \node[groupB, fit=(o1)(o5)(mltitle)] (mlgroup) {};
    \end{scope}
    
    \draw[arr] (goesflr.east) -- (arc.west);
    \draw[arr] (arc.east)     -- (fm.west);
    \draw[arr] (ssw.east)     -| ([xshift=-5mm]fm.south);
    \draw[arr] (xrt.east)     -| ([xshift= 5mm]fm.south);
    \draw[arr] (srs.east)     -- (srsc.west);
    \draw[arr] (srsc.north)   -- (arc.south);
    \draw[arr] (fm.east)      -- (dv.west);
    \draw[arr] (dv.east)      -- (sla.west);
    \draw[arr] (sla.south)    -- (docc.north);
    
    \draw[arr] (sharps.east) -- (mfp.west);
    \draw[arr] (mfp.east)    -- (aia.west);
    \draw[arr] (aia.east)    -- (fhp.west);
    \draw[arr] (aiad.east) -| (aia.south);
    \draw[arr] (doca.south)  -- (fhp.north);
    
    \draw[arr] (fhp.east)  -| (mvts.north);
    \draw[arr] (mvts.west) -- (mvts.west -| mlgroup.east);
    
    \end{tikzpicture}
    \caption{The block diagram of our dataset generation process, with principal procedures of flare cleaning (in red), MVTS generation and flare integration (in blue), and the eventual machine-learning-ready dataset creation (in orange).}
    \label{fig:pipeline}
\end{figure}
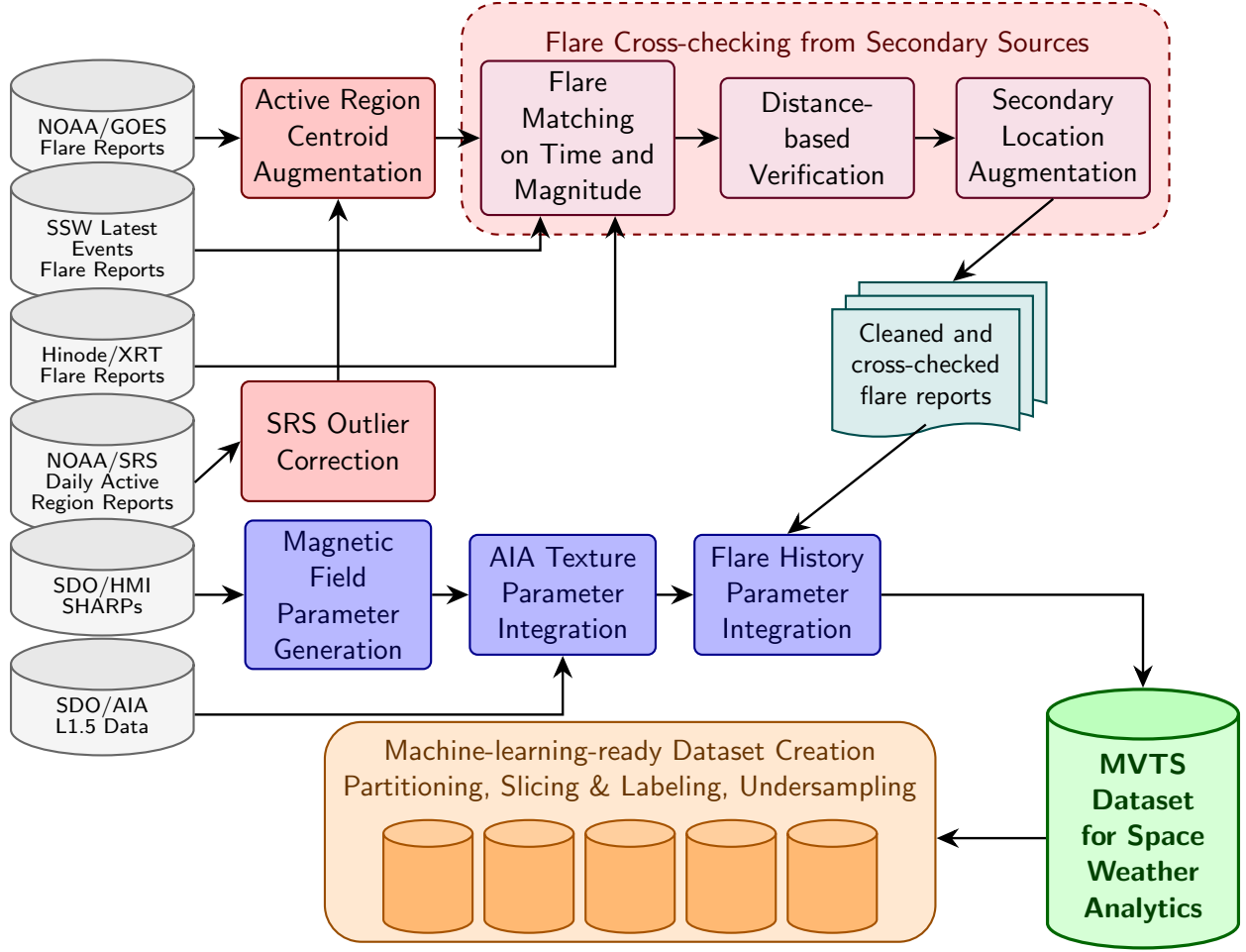

A schematic overview of the overall MVTS dataset generation process is presented in Figure~\ref{fig:pipeline}.

\section{Solar Flare Reports} \label{sec:flareinteg}
\begin{figure} [ht!]
  \centering
    \includegraphics[width=0.95\textwidth]{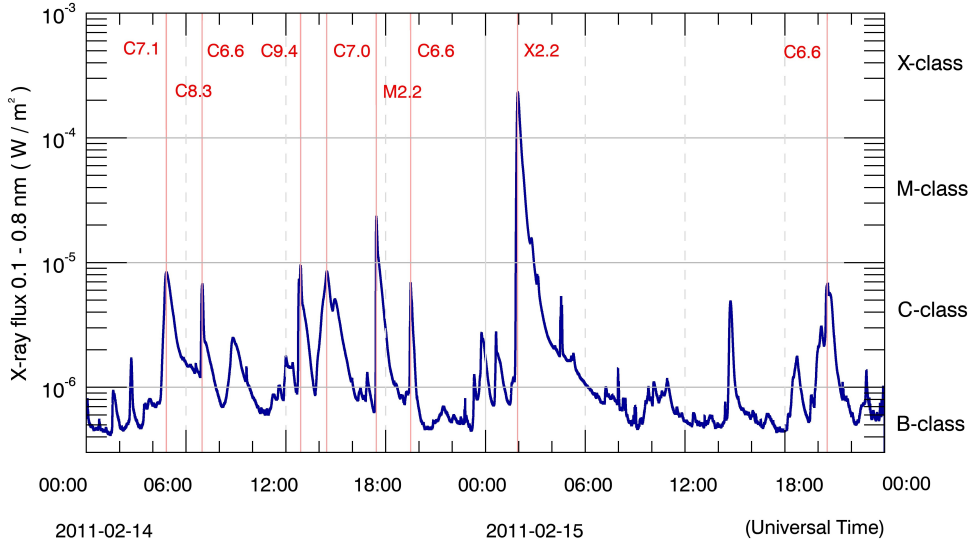}

  \caption{GOES15 1 - 8$\;$\AA$\;$solar X-ray flux from 2011-02-14 to 2011-02-15. The GOES flare classification is provided on the minor y-axis. The plot also includes annotations of flares exceeding GOES class C5.0, with red vertical lines indicating the peak times of the flares. The example interval also shows that during these two days of intense activity, background X-ray flux was high, making it difficult to identify small flares.}
  \label{fig:xray}
\end{figure}

In keeping with the original SWAN-SF dataset construction, we continue to consider the GOES flare catalog as our primary data source. The catalog utilizes NOAA/GOES observations \cite{hanser1996design} that measure disk-integrated fluxes between 0.1-0.8\,nm from the Sun using the X-ray Sensors (XRS). When a sudden, yet persistent, X-ray flux increase is detected, the event is flagged as a likely flare. Manual review is performed by NOAA forecasters to produce the final NOAA flare list. Some notes about the catalog are: (1) The GOES satellites are subject to eclipses by the Earth in the spring and fall, leading to interruptions (blackouts) in the X-ray flux record lasting from minutes to one hour; (2) The background X-ray radiation emitted by the Sun is usually at the level of A- or B-class flares (except certain periods during the solar maxima when it can reach the C-class flare level), making it difficult to capture all flares of these classes during higher-activity phases of the solar cycle. The M- and X-class flares (and most of the C-class flares), on the other hand, are seldom missed, except in periods of intense activity, when the background may even exceed C1.0. In Fig.~\ref{fig:xray} we present an example GOES X-ray flux series annotated with some flare occurrences; (3) As data from the XRS has no spatial information, NOAA uses data from the Solar X-ray Imager (SXI) on the same GOES satellites \cite{sxi2012}, which captures full-disk images with one-minute cadence in filter bands ranging from 0.6 to 6 nm, as well as other data sources, to pin-point each flare location. 

Utilizing the above spatiotemporal information about solar flares, NOAA's Space Weather Prediction Center (SWPC) then co-locates the active region responsible for the flare. Nonetheless, the GOES catalog is not perfect: the locations and NOAA active region numbers are missing for many B-, C- and even a few M-class flares. Our goal is to create a set of clean, cross-checked flare reports; therefore, we integrated the centroid locations of NOAA ARs to GOES flare reports without an explicit spatial location (i.e., only the NOAA AR numbers are listed), and later cross-checked these locations with two independent feature reporting modules, SSW Latest Events {\cite{SSWonline}} and Hinode-XRT {\cite{Watanabe2012}}. Hereafter, we will refer to flare reports from SSW Latest Events and Hinode-XRT catalog as SSW and XRT flares.

\subsection{Data Acquisition:}
We utilize the datasource classes produced for SWDataToolkit \cite{kempton:2026:swdatatoolkit} to curate the GOES , SSW, and XRT flares catalogs, along with NOAA AR locations from their Solar Region Summaries \cite{swpc:arsummary} (SRS) to enhance, verify, and clean the data. The GOES flare reports were are obtained from the Heliophysics Event Knowledgebase (HEK)\cite{hurlburt:hek}. The SSW flares were downloaded from their web archive {\cite{SSWonline}}, due to the inconsistencies between the web archive and records in HEK. The XRT flares were downloaded directly from the online XRT Flare Catalog {\cite{Watanabe2012}}.

During the period of interest that spans more than fifteen years (2010-05-01 to 2025-12-23), there are 30,061 GOES flare records, broken down to 141 X-, 2,482 M-, 18,326 C- and 9,014 B-, and 97 A-class events. We also downloaded 28,192 XRT flares and 35,627 SSW flares. All three data sources have the following common attributes: start time, peak time, end time, NOAA active region number, GOES class, and point location (i.e., heliographic latitude and longitude, in degrees).  Additionally, we utilize the daily NOAA active region list, which includes both numbered active regions with sunspots and plages, totaling some 29,367 daily NOAA active region reports.

\subsection{NOAA Active Region Database Cleaning}
Though not playing as central of a role in this version of the dataset as it did in  \cite{angryk2020multivariate, angryk2020multivariatedata}, the list of NOAA reported active regions is still a fundamental component of our data integration methodology. The NOAA active region (AR) locations are utilized as a means to augment unknown or not reported flare locations, in a process that is described in the next section. However, as was mentioned in the previous dataset release, it was identified that there are a number of instances where NOAA AR daily reports have unexpected location changes. The process used by \cite{angryk2020multivariate} to identify unexpected NOAA AR location reports was described in \cite{Cai2019}, and was largely focused on ensuring the association of the correct NOAA AR to the correct HARP, because it was using this association as the primary key in joining flare reports to HARP reports in the final timeseries dataset. In that work, the active regions were generally found to move between 13$^\circ$ to 14$^\circ$ per 24 hour report period. It was also found that some active regions moved towards east, did not change location, or moved over 25$^\circ$. Similarly, it was also observed that outliers existed for latitudes as well, where the active region latitude changed over 5$^\circ$ on a single day.

Utilizing these findings, we have used the algorithm outlined as Algorithm~\ref{alg:srs-outlier} to identify and correct outliers in the SRS AR reports. In this algorithm, we have set $\Delta\phi_{\max} = 5$, limiting the latitude displacement of consecutive daily SRS AR reports to 5$^\circ$.  We have selected $\Delta\lambda_{\min} = 5$ and $\Delta\lambda_{\max} = 25$, ensuring that longitude displacement of consecutive daily SRS AR reports to fall between 5$^\circ$ and 25$^\circ$.  In all, we identified 197 of the 29,367 daily NOAA active region reports as requiring correction under these settings. These updated locations are then stored and utilized for augmenting GOES flares with missing location information, but containing an associated NOAA AR number in their report information.  

%

\begin{algorithm}[t]
\caption{SRS active region location outlier detection and correction}
\label{alg:srs-outlier}
\begin{algorithmic}[1]
\Require Daily SRS reports $R$, each a tuple $(a, t, \phi, \lambda)$ of active region,
         date, latitude and longitude
\Statex \hspace{\algorithmicindent} Thresholds $\Delta\phi_{\max}$, $\Delta\lambda_{\min}$, $\Delta\lambda_{\max}$;
        visible-disk range $[\lambda_{lo}, \lambda_{hi}]$; minimum correction $\delta$
\Ensure  Corrected $(\phi, \lambda)$ and an outlier flag for every implausible report

\Statex
\State missing $\phi$ and $\lambda$ $\gets 180^\circ$
  \Comment{unusable values are forced to be flagged}
\ForAll{active regions $a$, with reports $R_a$ ordered by date}
  \State $O_a \gets \emptyset$ \Comment{outlier dates for region $a$}
  \ForAll{consecutive pairs $(r_{i-1}, r_i)$ in $R_a$ one day apart}
    \If{$|\phi_i - \phi_{i-1}| > \Delta\phi_{\max}$}
      \State $O_a \gets O_a \cup \{t_i\}$ \Comment{latitude not allowed to drift this fast}
    \ElsIf{$|\lambda_i - \lambda_{i-1}| \notin [\Delta\lambda_{\min}, \Delta\lambda_{\max}]$}
      \State $O_a \gets O_a \cup \{t_i\}$ \Comment{inconsistent with solar rotation}
    \EndIf
  \EndFor

  \ForAll{$\tau \in O_a$ in chronological order}
    \State $p \gets$ latest report in $R_a$ before $\tau$;\quad
           $n \gets$ earliest report in $R_a$ after $\tau$
    \If{$p$ and $n$ both exist}
      \State $(\phi', \lambda') \gets$ linear interpolation of $(\phi, \lambda)$ between $p$ and $n$ at $\tau$
    \Else
      \State $m \gets$ whichever of $p$, $n$ exists
      \State $(\phi', \lambda') \gets \Call{RotateSun}{\phi_m, \lambda_m, t_m, \tau - t_m}$
        \Comment{differential rotation; signed offset}
    \EndIf
    
      \State write $(\phi', \lambda')$ and flag $\tau$ as an outlier
  \EndFor
\EndFor
\end{algorithmic}
\end{algorithm}

\subsection{Data Enhancement and Verification for GOES Flares:}
\label{sec:flare_verification}
As was done with the original dataset, the first step of the solar flare enhancement and cross-checking procedure involves a data enrichment process for GOES flares that are missing an explicit point location, but have an associated NOAA AR. Additionally, for GOES flare reports that do have location information, as well as NOAA AR associations, we utilize the temporally nearest SRS report to determine where the associated active region would be at the time of the flare. We utilize the known solar differential rotation to rotate the reported active region location to that point in time, and test if the flare reported location is greater than 7$^\circ$ degrees from the rotated AR location.  If, so, the GOES flare is augmented with the rotated AR location to better match the likely HARP location used in HARP association later.  

%

\begin{algorithm}[ht!]
\caption{Merging GOES, SSW and XRT flare reports, HARP assignment and confidence}
\label{alg:flare-validation}
\begin{algorithmic}[1]
\Require Flare reports $G$ (GOES), $S$ (SSW), $X$ (XRT), each $(t, c, \phi, \lambda)$ of
         peak time, GOES class, latitude and longitude; HARP records $\mathcal{H}$
\Statex \hspace{\algorithmicindent} Peak-time tolerance $\Delta t_{\max}$; class flux ratio $\rho_{\max}$;
        agreement distance $d$ ($4^\circ$); 
\Statex \hspace{\algorithmicindent} HARP buffer $b$
\Ensure  Verified flares, each with a HARP number and a confidence tier

\Statex
\Function{Match}{$P, Q$} \Comment{global one-to-one assignment}
  \State $E \gets \{(p, q) : |t_p - t_q| \le \Delta t_{\max},\ f_{\max}/f_{\min} \le \rho_{\max}\}$
    \Comment{$f$: flux}
  \State accept $E$ greedily by $(|t_p - t_q|, |\log f_p/f_q|)$; each $p, q$ used once
\EndFunction

\Statex
\State $F \gets G \cup \Call{Match}{G, S} \cup$ non-duplicate unmatched $S$
\State $F \gets F \cup \Call{Match}{F, X} \cup$ non-duplicate unmatched $X$

\ForAll{merged events $e \in F$}
  \State $L \gets \{s \mapsto (\phi_s, \lambda_s) : s \in \{G, S, X\}$ has a usable location$\}$
  \ForAll{$s \in L$} \Comment{buffered HARP containment at $t_e$}
    \State $H_s \gets \Call{HARPs}{t_e, \overline{L}, b}$ \textbf{if} all of $L$ within $d$
            \textbf{else} $\Call{HARPs}{t_e, L_s, b}$
    \State $h_s \gets$ member of $H_s$ whose box contains $L_s$, else nearest center
  \EndFor
  \State $h^* \gets$ the HARP chosen by a majority of the sources with an $h_s$
  \State $a \gets$ locations of sources supporting $h^*$ are pairwise within $d$
  \State $\text{tier} \gets \begin{cases}
      \text{unverified} & \text{no } h_s \\
      \text{very\_high} / \text{high} & 3\text{-}0,\ a \,/\, \neg a \\
      \text{high} / \text{moderate} & 2\text{-}0,\ a \,/\, \neg a \\
      \text{low} & 2\text{-}1 \text{ or } 1\text{-}0 \\
      \text{very\_low} & \text{all } h_s \text{ distinct (1-1, 1-1-1)}
    \end{cases}$
  \If{tier $=$ very\_low}
    \State emit one row per source $s$ with $(h_s, L_s)$
  \Else
    \State emit one row with $h^*$ and the mean location of its supporters
  \EndIf
\EndFor
\end{algorithmic}
\end{algorithm}

Among 30,061 GOES flares, only 6,690 have explicit locations and 23,332 do not. For these missing locations, we use the associated NOAA active region locations as a proxy. With NOAA active region location augmentation, we determined the approximate locations of 12,550 flares.  The vast majority of the remaining 10,782 GOES flares with undetermined locations were A-, B- and C-class flares and 9,409 or 87.27\% of them did not have location information or NOAA active region association. 

Once the GOES flare reports are augmented, we then attempt to match each GOES flare to an SSW and an XRT flare using the temporal attributes (start and end times of flares) and flare magnitudes. In Algorithm~\ref{alg:flare-validation}, we outline the procedure used to perform flare cross-matching and assignment to HARP reports as this procedure differes from that used in \cite{angryk2020multivariate, angryk2020multivariatedata} for the original dataset. In this procedure, we pair reports whose peak times are within $\Delta t_{\max} = 600$ seconds and whose class fluxes are within a ratio of $\rho_{\max} = 5.0$ of each other. It then accepts pairs greedily, closest peak time first and class difference second, using each report at most once. We start with GOES reports, pairing them with SSW (GOES + SSW), and the combined events are then paired with XRT. Any unmatched SSW and XRT reports are later added as standalone events, unless they duplicate one that was already matched. In this matching and merging, the location is not used to decide which reports are the same flare.

Once the list of cross-matched flare reports is constructed, the reports are then assigned to HARP reports by collecting usable locations, and if they all agree within $4^\circ$, we query at their mean location at the GOES reported peak time; otherwise each flare report location is separately queried. For each report, the HARP whose box contains the location is chosen, or the HARP with the nearest center is chosen if no bounding box contains the queried location. The assignment of confidence tier comes from the HARP vote: 3-0 gives ``very high'' or ``high'', 2-0 gives ``high'' or ``moderate' (depending on whether the locations are within $4^\circ$), 2-1 or 1-0 gives ``low'', and if all votes are different then ``very low'' is assigned and one row per source is generated. When no HARP is found, the report is labeled as ``unverified''. With the exception of the ``very low'' case, every event produces one row for the winning HARP.

The reported locations from GOES, SSW, and XRT are usually different. While most of these differences are negligible, some are not. There are a variety of reasonable explanations for these differences, including the accuracy of the reported coordinate (i.e., the number of decimal places reported), use of approximate active region location augmentation (both by us and by XRT), or pixel bleeding \cite{Poduval2013}. However, for the large differences, often times reporting modules either do not report correct coordinates (such as small flares close to poles) or there are multiple flares occurring close to the solar limb.

 \begin{figure} [b!]
   \centering
     \begin{tabular}{c}
 	  	\includegraphics[width=0.84\textwidth]{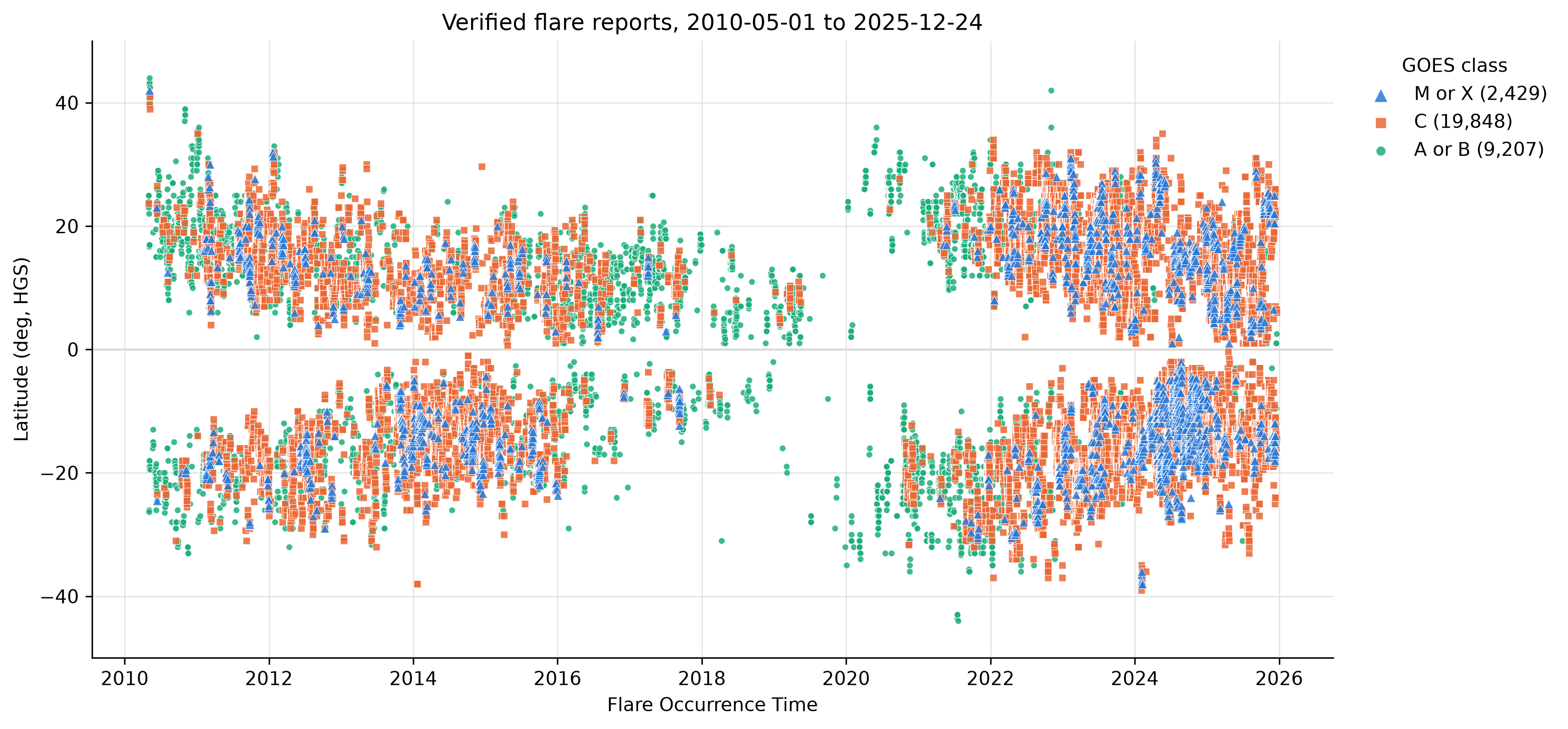} \\
   \end{tabular} \\
   \caption{Scatter plot of the verified heliographic latitudes of flares (in degrees), as a function of peak times, ranging between May 1, 2010 and December 24, 2025.
   }
   \label{fig:flareresults}
 \end{figure}

\subsection{Resulting Flares}

After the NOAA AR augmentation and flare cross-checking steps, between May 1 2010 and December 24, 2025 we have 7,816 very high confidence, 12,825 high confidence, 1,401 moderate confidence, 9,443 low confidence, 797 very low confidence, and 7253 non-verified GOES flares. There are 125 X-class, 2304 M-class, 19,848 C-class, and 9,207 A- or B-class flares that were verified and not in the very low confidence group. There are 28 of 153 X-class flares that are unverified. Only 346 out of 2702 M-class (12.8\%) are not verified. About 83.7\% of C-class flares and 76.3\% of A- and B-class flares are verified. The majority of non-verified flares are, therefore, A- and B-class events. Fig.~\ref{fig:flareresults} depicts the scatter plot of verified flare latitudes and peak times as a function of time, which is reminiscent of the long known butterfly diagram for sunspots or solar ARs \cite{Hathaway2015}.

\section{SHARP Data and Magnetic Field Parameters}\label{sec:cleanmvts}

A HARP data collection (and the corresponding SHARPs) consist of a 12-minute sampled time series of spatial cutouts including the vector magnetic field, continuum intensity, and maps or values of other quantities. Each HARP may contain one or more solar active regions within the cutout region. Each HARP series is labeled with a unique identifier, \texttt{HARPNUM}. The number of observations in HARP series depends on how long the active region(s) it encloses were visible on the solar disk.

There are two types of HARPs (and associated SHARP metadata) available from JSOC: the \emph{definitive} and the \emph{near real-time} (\emph{NRT}). The NRT series is useful for space weather forecasting in an operational context as it is processed within three hours of acquisition. However, the NRT dataset pipeline changes the bounding box size of HARPs as they evolve and assigns different identifiers to active regions within the series that might merge or split as they traverse the disk. This makes it difficult to associate flares to specific HARPs. As was the case in the original dataset of \cite{angryk2020multivariate, angryk2020multivariatedata}, this is why we have chosen to utilize the definitive series instead.

The definitive series is processed after observing a HARP for its entire rotation across the earthward solar hemisphere. A maximal bounding box, which can often encompasses multiple active regions within a HARP, is chosen and remains fixed in this case. Active regions that merge or split are also tracked as a single, all-encompassing HARP. The higher data quality and consistency makes the definitive series a better option for creating benchmark datasets that increase our physical understanding of space weather phenomena and their possible links to the photospheric magnetic field, including the identification and optimization of solar flare predictors. 

HARP magnetogram time series are available in two coordinate systems: native CCD and Lambert cylindrical equal area (CEA, {\cite{Thompson2006}}). In the CEA projection, the vector magnetic field is decomposed into radial ($r$), westward ($\phi$), and southward ($\theta$) components. This projection is very convenient for calculating various extensive (i.e., area- or size-dependent) quantities, such as the total area of the active region, its magnetic flux, etc. For our dataset, we have used the raster data in the definitive series mapped to CEA projection with 720 seconds cadence (\verb|hmi.Sharp_cea_720s|).

\subsection{Magnetic Field Parameters:} 
\def\mathbi#1{\textbf{\em #1}}

\begin{table}[ht!]
\caption{Computed magnetic field parameters. Parameters marked with asteriks (*) are discussed in \cite{bobra2015solar}, but are not available in SHARP headers. }
\label {tab:mfp}

\resizebox{.99\textwidth}{!}{%

\renewcommand{\arraystretch}{1.4}
\begin{tabular}{  r  l  l  }

Magnetic Field  & &  \\

 Parameters from \cite{bobra2015solar} & Description & Formula \\ 

 \hline &    \\ [-0.5cm] 
 
{ABSNJZH} \cite{Leka2003} 	& Absolute value of the net current helicity in G2/m   & $H_{c_{abs}} \propto | \sum B_z \cdot J_z | $  \\ [0.1cm]
 
{EPSX*}	\cite{Fisher2012} & Sum of X-component of normalized Lorentz force & $\delta F_x \propto \frac{\sum B_x B_z}{\sum B^2}$ \\ [0.1cm]

{EPSY*}	\cite{Fisher2012} & Sum of Y-component of normalized Lorentz force & $\delta F_y \propto \frac{-\sum B_y B_z}{\sum B^2}$   \\  [0.1cm]

{EPSZ*} \cite{Fisher2012}	& Sum of Z-component of normalized Lorentz force & $\delta F_z \propto \frac{\sum(B_x^2 + B_y^2 - B_z^2)}{\sum B^2}$ \\ [0.1cm]

{MEANALP} \cite{Leka1999}	& Mean twist parameter & $\alpha_{total} \propto \frac{\sum J_z \cdot B_z}{ \sum B_z^2}$ \\ [0.1cm]

{MEANGAM} \cite{Leka2003}	& Mean inclination angle & $\bar{\gamma} = \frac{1}{N} \sum \arctan \left( \frac{B_h}{B_z}\right)$ \\  [0.1cm]
 
{MEANGBH} \cite{Leka2003}	&  Mean value of the horizontal field  gradient  & $\overline{\nabla B_h} =  \frac{1}{N} \sum \sqrt{ (\frac{\partial B_h}{\partial x} + \frac{\partial B_h}{\partial y} ) }$   \\ [0.1cm]

{MEANGBT} \cite{Leka2003}	& Mean value of the total field  gradient & $\overline{|\nabla B_{tot}|} = \frac{1}{N} \sum \sqrt{ ( \frac{\partial B}{\partial x} + \frac{\partial B}{\partial y} ) } $\\ [0.1cm]

 {MEANGBZ} \cite{Leka2003}	& Mean value of the vertical field gradient  & $\overline{\nabla B_z} =  \frac{1}{N} \sum \sqrt{ ( \frac{\partial B_z} {\partial x} + \frac{\partial B_z}{\partial y} ) }$  \\ [0.1cm]

{MEANJZD} \cite{Leka2003}	& Mean vertical current density & $\overline{J_z} \propto \frac{1}{N} \sum ( \frac{\partial B_y}{\partial x} - \frac{\partial B_x}{\partial y} )$ \\ [0.1cm]

{MEANJZH} \cite{Leka2003}	& Mean current helicity & $\overline{H_c} \propto \frac{1}{N} \sum B_z \cdot J_z$ \\  [0.1cm]

{MEANPOT}	\cite{Wang1996}	& Mean photospheric excess magnetic energy density & $\bar{\rho} \propto \frac{1}{N} \sum \left(\vec{\mathbi{B}}^{Obs} -  \vec{\mathbi{B}}^{Pot}  \right)^2 $ \\ [0.1cm]
  
{MEANSHR} 	\cite{Wang1996}	& Mean shear angle  & $\bar{\Gamma} = \frac{1}{N} \sum \arccos \left( \frac{ \vec{\mathbi{B}}^{Obs} \cdot \vec{\mathbi{B}}^{Pot} }{ |B^{Obs}| |B^{Pot}| } \right)$ \\ [0.1cm] 
 
{R\_VALUE*}	\cite{Scrijver2007}	& \begin{tabular}{@{}l} Total unsigned flux around high gradient \\ polarity inversion lines using the $B_r$ component \end{tabular} & $\Phi=\Sigma |B_{r}|.dA~(within~R~mask)$ \\ [0.1cm]
   
{SAVNCPP} \cite{Leka2003}	& \begin{tabular}{@{}l} Sum of the absolute value of the net current \\ per polarity \end{tabular} & $J_{z_{sum}} \propto \left| \sum\limits^{B_z^+} J_z dA \right| + \left| \sum\limits^{B_z^-} J_z dA \right| $\\  [0.1cm]
 
{SHRGT45} \cite{Leka2003}	& Area with shear angle greater than 45 degrees & $\frac{ \mathsf{Area~with~Shear} >45^{\circ} }{ \mathsf{Total~Area} } $\\  [0.1cm]
 
{TOTBSQ*} \cite{Fisher2012}	& Total magnitude of Lorentz force & $F \propto \sum B^2$  \\ [0.1cm] 
 
{TOTFX*} \cite{Fisher2012} & Sum of X-component of Lorentz force & $F_x \propto \sum B_x B_z dA$ \\ [0.1cm]
 
{TOTFY*} \cite{Fisher2012} & Sum of Y-component of Lorentz force & $ F_y \propto \sum B_y B_z dA $  \\ [0.1cm]
 
{TOTFZ*} \cite{Fisher2012} & Sum of Z-component of Lorentz force & $F_z \propto \sum(B_x^2 + B_y^2 - B_z^2)dA$ \\ [0.1cm]

{TOTPOT} \cite{Leka2003}	& Total photospheric magnetic energy density   & $\rho_{tot} \propto \sum \left(\vec{\mathbi{B}}^{Obs} - \vec{\mathbi{B}}^{Pot} \right)^2 dA$ \\ [0.1cm]

{TOTUSJH} \cite{Leka2003}	& Total unsigned current helicity & $H_{c_{total}} \propto \sum | B_z \cdot J_z | $  \\ [0.1cm]

{TOTUSJZ} \cite{Leka2003}	& Total unsigned vertical current & $J_{z_{total}} = \sum |J_z|dA$ \\ [0.1cm] 
 
{USFLUX} \cite{Leka2003}	& Total unsigned flux in Maxwells & $\Phi = \sum |B_z|dA$ \\ [0.1cm]
 
 \hline
\end{tabular} }

\end{table}
It has become generally accepted that, since flares are predominantly magnetic phenomena, a viable flare forecast could rely on the choice of adequate magnetic field properties and prediction methods (see, e.g., \cite{cui2007correlation, Georgoulis2012, ahmed2013solar, bobra2015solar, Barnes2016}). Therefore, we use the definitive \verb|hmi.Sharp_cea_720s| data series to calculate the parameters discussed in \cite{bobra2015solar} using the vector magnetic field. We have chosen to recalculate these parameters ourselves, first, to validate these data and achieve better maintainability and, second, to complement them with parameters not currently present in SHARP headers but considered important for flare and coronal mass ejection prediction. An example of such parameters are the shear angle weighted SHARP parameters discussed in Section~\ref{section:weighted_sharps}.

Prior to the release of the original SWAN-SF dataset of \cite{angryk2020multivariate, angryk2020multivariatedata}, many, but by no means all, of the flare-prediction studies did not consider these magnetic field parameters as time series. Instead, forecasting relied on cross-sectional, or point-in-time (snapshot) parameter values\cite{Barnes2016, Leka2019, Leka2019b}. There were a few exceptions: Gallagher et al.\ \cite{GallagherMW02}, Falconer et al.\ \cite{FalconerMBK2014}, and Leka et al.\ \cite{LekaBW2018} used the flaring rate or previous flaring in an active region. However, after the release of the original dataset, a change in approach was enabled, where many papers on training models on time series data in cases of extreme class imbalance were studied \cite{Ahmadzadeh2019, AhmadzadehICMLA2019, Hostetter2019, Chen2019-MVTSprec, chen2020, Ji2020, Ahmadzadeh2021how, Chen2021, Ma2019solar, chen2022cgan, wen2022improving, ji:2022:dnn_flare} to list a few. It is our intent to enable further such investigations through the release of additional parameters, and extending the coverage period of the dataset. Therefore, as was done previously, to facilitate both point-in-time and time series analysis, we derive a set of magnetic field parameters from individual region patches and transform them into multivariate time series over the entire length of a given HARP series. 

A number of physically important and potentially flare-predictive magnetic field parameters have been listed by \cite{bobra2015solar} and are reproduced in Table~\ref{tab:mfp}. However, as previously mentioned, several parameters (marked with an asterisk in Table~\ref{tab:mfp}) are not included in the SHARP header information but are present in our MVTS. For the generation of these parameters, we used the following information: $B_r$ (radial component of the magnetic field), $B_r error$ (the per-pixel uncertainty of the radial component of the magnetic field), $B_\theta$ (southward/azimuthal component of the magnetic field), $B_\theta error$ (the per-pixel uncertainty of the southward/azimuthal component of the magnetic field), $B_\phi$ (westward/toroidal component of the magnetic field), $B_\phi error$ (the per-pixel uncertainty of the westword/toridal component of the magnetic field), \emph{BITMAP} (active region boundary), \emph{MAGNETOGRAM} (line of sight magnetogram), and \emph{CONF\_DISAMB} (confidence map of magnetic field disambiguation). The previous iteration of the dataset utilized these segments as inputs to a GPU implementation of the magnetic field parameter calculation module \cite{BasodiAA17}. This module, however, suffered from multiple shortcomings, most notably the exclusion of the uncertainty calculations for many of the magnetic field parameters and it's poor documentation. So, as part of development for this dataset, a CPU based library was developed to contain all of the relevant data sourcing and parameter calculation modules. A description of this Space Weather Toolkit Library can be found in \cite{kempton:2026:swdatatoolkit}. Note that as part of the development of the Space Weather Toolkit Library, magnetic parameter calculations were extensively tested against SHARP keyword values to verify the recalculated parameters correctness.

Utilizing this library, we generated the time series of all magnetic field parameters listed in Table~\ref{tab:mfp}. However, the derived parameters each have uncertainty inherent in their estimate due to the uncertainty in the magnetic field estimates at each pixel included in their calculation. So, we have elected to include many of their propagated uncertainties as part of this release of the dataset. The inclusion of the uncertainties is a new addition for this dataset, and due to that, many previous prediction methods based on these parameters have neglected this inherent uncertainty in the outputs.  However, since the release of the original dataset, NASA's Space Radiation Analysis Group (SRAG), the entity responsible for ensuring that the radiation exposure received by astronauts remains below established safety limits, has developed a set of requirements for any monitoring and modeling tools developed to support their decision making \cite{srag-model-requirements}. In this document, SRAG outlines several integrated space weather project requirements, from probability of event occurrence forecasts, to forecasts about event onset time and the intensity profiles of events that are expected to occur. One key discussion in this document revolves around a description of the expected functionality of ``all-clear'' models as needing to include model uncertainty in each of the forecasts. This is the primary motivation for our inclusion of the parameter uncertainties in this release.

\subsection{Weighted Magnetic Field Parameters}\label{section:weighted_sharps} 
There have been numerous studies showing that solar active regions (ARs) that produce major flares typically exhibit strong plasma shear flows around photospheric magnetic polarity inversion lines (MPILs)\cite{park:2018, Georgoulis:2012:non-neutralized, Tziotziou:2013}. In \cite{Tziotziou:2013}, the authors state that a physically meaningful way to quantify the non-potentiality or, in other words, the eruptive capacity of solar active regions (ARs) is by calculating these regions’ non-potential, ``free'' magnetic energy. This free magnetic energy is the excess energy stored in the magnetic configuration of the AR, above the reference, potential energy corresponding to the AR's magnetic energy in the absence of electric currents. Here magnetic helicity, is the corresponding degree of twist, writhe, and linkage of the AR's magnetic field lines, and a nonzero (left- or right-handed) total magnetic helicity implies nonzero free energy\cite{Berger:1999}. 

Methods used to calculate such shear/helicity use the deviation of an observed vector magnetic field from the potential magnetic configuration. This is referred to as the magnetic non-potentiality, and many non-potential parameters have been introduced to measure the magnetic non-potentiality of ARs. In these parameters, the electric current density is directly calculated from the curl of a vector magnetic field. However, due to a lack of accurate observations of the magnetic field in the higher layers of the solar atmosphere, the vertical component of electric current density ($J_z$) in the photosphere is widely used \cite{Song:2013}. The calculation of angular shear is then derived from the differences between the observed magnetic field and the potential magnetic field, which is extrapolated from the observed photospheric field with assumptions about the potential field. In this definition, the angular magnetic shear is considered the angular degree difference between the observed horizontal field and the horizontal component of the potential field \cite{hagyard:1984}. The lack of electric currents automatically means a zero free magnetic energy and a zero relative helicity, at the same time meaning the inability of an AR to energetically power solar eruptions, that is, flares and coronal mass ejections. So, we take these studies as motivation for utilizing the magnitude of the sine of the angular shear to focus calculations of magnetic parameters to only those regions that have the high potential to influence the eruptive potential of the AR.

\begin{figure}[htb!]
    \centering
    \includegraphics[width=0.95\textwidth]{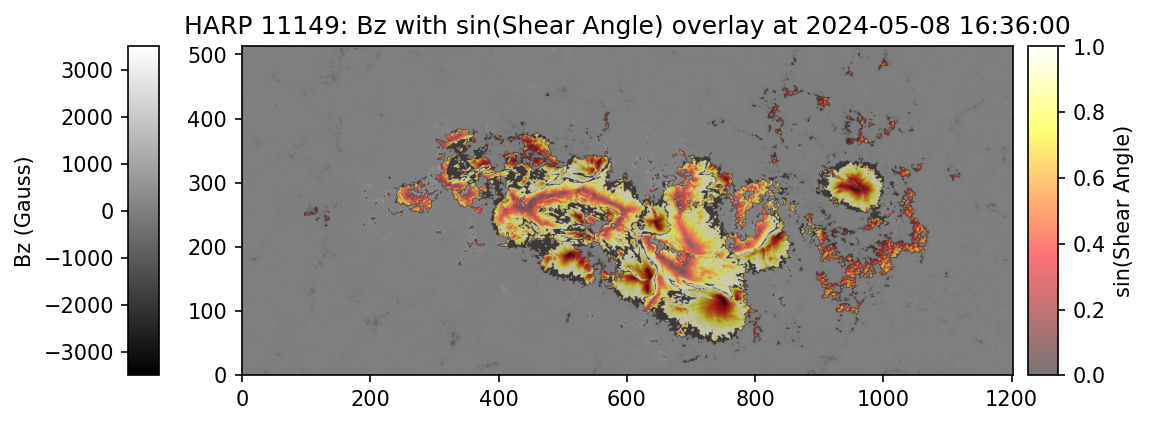}
    \caption{The $B_z$ parameter for HARP 11149 with the $sin(shear\ angle)$ overlay applied to areas where $|B_z| > 100G$.}
    \label{fig:harp_w_shear_visual}
\end{figure}

In Figure~\ref{fig:harp_w_shear_visual}, we show the $sin(shear\ angle)$ applied as a weighting mask to $B_r$ (radial component of the magnetic field) component of HARP 11149.  In \cite{Georgoulis:2012:non-neutralized}, a process is described by which nonzero azimuthal Lorentz forces will cause shearing motions along the PIL and give rise to magnetic shear, and in the figure we do indeed observer a strong shear on both sides of the PIL. We apply this weighting mask to every HARP that we calculate the above listed magnetic parameters for, giving us a weighted set of parameters that are highly correlated to the unweighted SHARP parameters listed in Table~\ref{tab:mfp}.

\subsection{Cleaning the MVTS} 

The dataset is composed of 7,730 distinct HARP regions with magnetic field files, even though the numbering of HARPs ranges from 1 to 14,172. The HARPS numbers in the definitive series that contain no reported data from JSOC are likely the result of merging active regions and other filtering steps of the definitive series data construction algorithm. Though the HARP numbers in these two systems are not direct a direct  The individual records of these HARPs result in over 990,000 hours of solar activity observations. Furthermore, about 1.09\% of timestamps were missing in the time series and these were filled with \emph{null} values to maintain the fixed cadence of 12 minutes. Potential reasons for these data gaps are, first, gaps in the SHARP series when the HARP is close to the eastern solar limb or when it is about to rotate beyond the western limb and, second, the SDO eclipse seasons.

\section{Augmentation Through Use of SDO/AIA Extreme Ultraviolet Images}\label{section:aia_params}
    
While related to the same active regions, the magnetic field and Ultraviolet / Extreme Ultraviolet (UV/EUV) parameters refer to the state and dynamics of different layers of the solar atmosphere (photosphere and chromosphere-corona, respectively) and, therefore, are yet expected to behave differently. Several works have previously identified flare precursors in the EUV image parameters and UV spectra about 1 hour before the flare occurrence \cite{Krista:2025,krista2021deftflares,panos2020IRISflareprediction,woods2021IRISflares}. Correspondingly, one may expect UV/EUV parameters to be of use for the flare predictions and complementary to existing magnetic field descriptors. This is our motivation for expanding the SWAN-SF benchmark dataset to take advantage of this wealth of high-dimensional data as well.

\begin{table}[ht!]
        \footnotesize
        \centering
        \caption{The ten image parameters computed on SDO/AIA observations for augmentation of SWAN-SF. In the formulas, $L$ denotes the maximum intensity value (e.g. $255$ in an 8-bit image (though we use real valued floats after AIA Fits L1.5 preprocessing)), $i$ denotes the color intensity value $(i \in [0, L])$, $p$ denotes the probability (i.e., normalized histogram), $h$ denotes the histogram, $N$ denotes the number of counting boxes, and $\varepsilon$ denotes the side-length of the counting box.}
        \vspace{0.5 cm}
        \begin{tabular}{r | l  l}
            \toprule
            & Image Parameter & Formula \\
            \toprule
            1 & Entropy & \( - \sum_{i=0}^{L}{p(i) \cdot \log_2(p(i))}\) \\ \midrule
            2 & Mean ($\mu$) &  \( \sum_{i=0}^{L}{h(i)\cdot i}\)\\ \midrule
            3 & Standard Deviation ($\sigma$) & \( \sqrt{\sum_{i=0}^{L}h(i) \cdot (i - \mu)} \) \\ \midrule
            4 & Fractal Dimension & \( - \lim\limits_{\varepsilon \to 0}{ \frac{\log(N)}{\log(\varepsilon)}} \)  \\ \midrule
            5 & Skewness ($\mu_{3}$) & \( {\frac{1}{\sigma^{3}}} {\sum_{i=0}^{L}{h(i)(i - \mu)^3}} \) \\ \midrule
            6 & Kurtosis ($\mu_{4}$) & \( {\frac{1}{\sigma^{4}}} {\sum_{i=0}^{L}{h(i)(i - \mu)^4}} \) \\ \midrule
            7 & Uniformity & \( \sum_{i=0}^{L}{p^2(i)}\) \\ \midrule
            8 & Relative Smoothness & \( 1 - {\frac{1}{1+ \sigma ^2}}\) \\ \midrule
            9 & Tamura Contrast &  \( \frac{\sigma ^2}{{\mu_4} ^ {0.25}} \) \\ \midrule
            10& Tamura Directionality & See Eq. 3 in \cite{Ahmadzadeh2019curated}\\ \midrule
            \bottomrule
        \end{tabular}
        \label{table:tenParams}
    \end{table}

The atmospheric imaging assembly (AIA) has four telescopes that provide narrow-band imaging of seven extreme ultraviolet (EUV) band passes ($94$\,{\AA}, $131$\,{\AA}, $171$\,{\AA}, $193$\,{\AA}, $211$\,{\AA}, $304$\,{\AA}, and $335$\,{\AA}) and two ultraviolet (UV) channels ($1600$\,{\AA} and $1700$\,{\AA}) \cite{lemen:sdoAIA}. The captured $4k$ images of the Sun, which are full-disk snapshots with the cadence of $12$ seconds, are compressed on board and without being recorded on orbit, are transmitted to SDO ground stations. The received raw data (Level 0) are archived on magnetic tapes in JSOC science-data processing facility. The uncompressed data is then exported as FITS files with the data represented as $32$-bit floating values. At this point, images are already calibrated, however, some corrections and cleaning are still required due to the existence of a small residual roll angle between the four AIA telescopes. At this stage (Level 1.5), the data is ready for analysis. In some repositories including Helioviewer, the FITS files are converted to JP2 format in order to reduce the volume of their database.

In \cite{Ahmadzadeh2019curated}, a set of ten image parameters (listed in Table~\ref{table:tenParams}) was extracted from the SDO/AIA's (EUV and UV) observations and investigated for a number of different tasks. These parameters are included as part of the Space Weather Toolkit Library \cite{kempton:2026:swdatatoolkit} and are utilized an additional augmenting set of parameters for the SWAN-SF dataset. These image parameters have already been successfully used for the classification of filaments in H-alpha images from the Big Bear Solar Observatory (BBSO) \cite{schuh2014massive}, and for building a Content-based Image Retrieval (CBIR) System \cite{schuh2015solar}. Later on, the same set of image parameters were used for building an automatic module, using sparse coding technique, for tracking solar filaments \cite{kempton2015tracking, kempton2016towards, kempton:phenomena}.

\begin{figure}[htb!]
    \centering
    \includegraphics[width=0.95\textwidth]{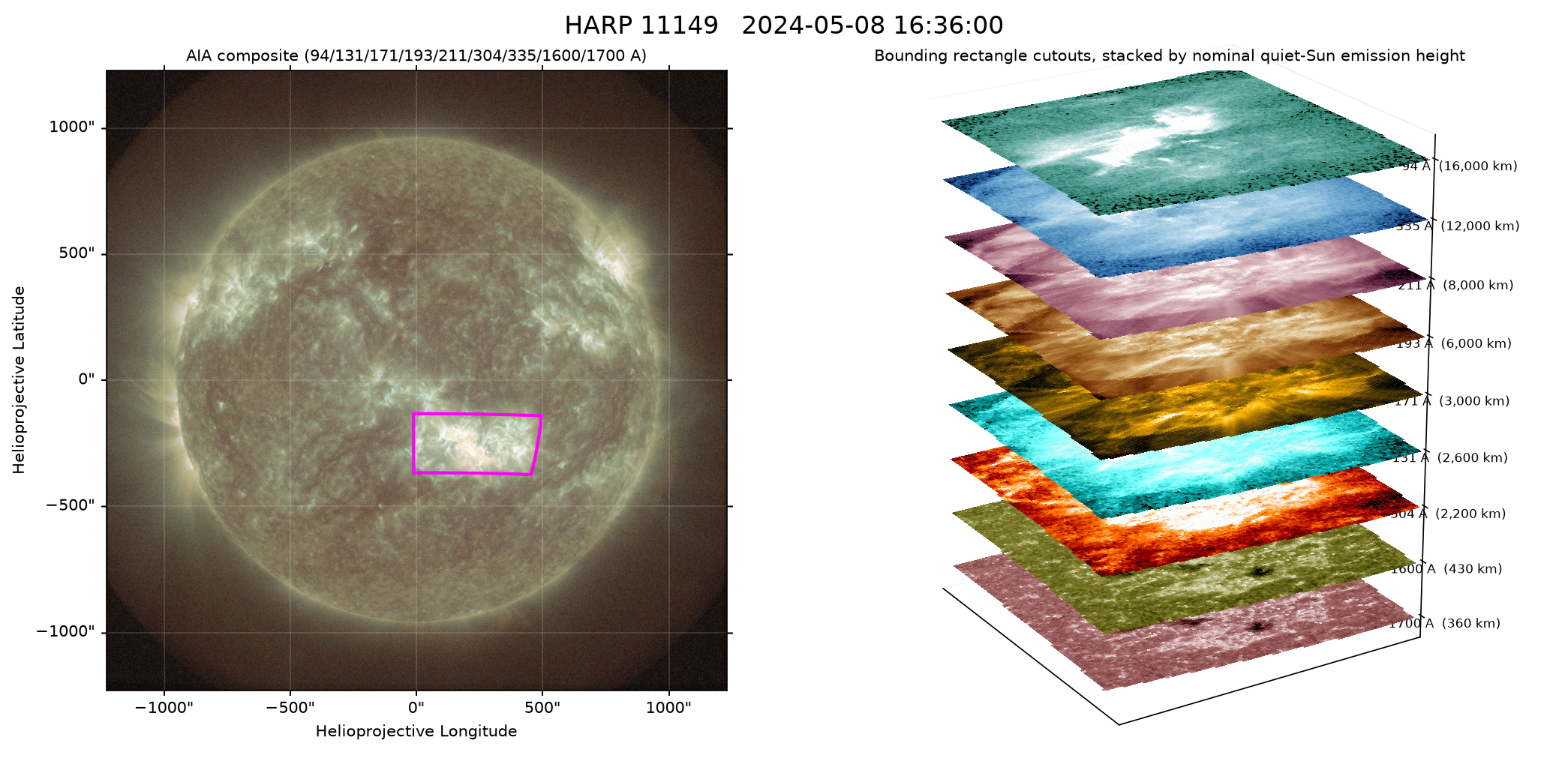}
    \caption{The resultant cutouts for HARP 11149.}
    \label{fig:cutout_visual}
\end{figure}

All parameters in Table~\ref{table:tenParams}, except for the fractal dimension and Tamura directionality, capture some information about the distribution of the pixel intensity values of the images. In \cite{Ahmadzadeh2019curated}, in order to preserve the spatial information of the data, a grid-based segmentation on the images was utilized. This is a widely used technique already experimented on the AIA images, which has shown good results \cite{banda2009effectiveness, banda2010experimental}. This way, each $4096\times4096$-pixel AIA image is segmented by a fixed $64\times64$-cell grid. For each grid cell that spans over a square of $64\times64$ pixels of the image, the $10$ image parameters will be calculated. Since $10$ parameters are processed for each image, the image then forms a data cube of size $64\times64\times10$. Additionally, for each time step, $9$ images are processed from different wavelength filter channels of the AIA instrument.

To augment the SWAN-SF benchmark data set, however, each observation of an active region (or more precisely, an active region's identified bounding box) should correspond to a vector of 10 image parameters which change with time. Since each observation captured by the SDO/AIA corresponds to 9 different channels within the invisible-light spectrum (EUV and UV), this will add 90 new variables to the existing multivariate time series of SWAN-SF. So, as a means to limit the number of variates added, no grid-based segmentation will be used for this purpose. Instead, we have calculated each of the 10 image parameters over the whole AIA cutout.  The process of computing a region of interest is described in Section~\ref{section:aia_cutout}, and experiments performed to find optimal settings for the texture parameters that have additional hyper parameters is described in Section~\ref{section:text_param_exp} below.

\subsection{AIA Cutout Processing}\label{section:aia_cutout}

\begin{figure}[htb!]
    \centering

    \noindent
    \begin{tikzpicture}[
      >={Stealth[length=2.3mm]},
      every node/.style={font=\small, text=ink},
      lbl/.style={font=\footnotesize},
      tny/.style={font=\scriptsize},
      guide/.style={densely dashed, gray!65, line width=0.35pt},
      trim left,
    ]
    
    \begin{scope}[xshift=3.7cm]
      \def\R{2.25}       
      \def\Rh{3.10}      
      \def\pa{50}        
    
      \draw[sunedge, densely dotted, line width=0.6pt] (0,0) circle (\Rh);
      \filldraw[fill=sunfill, draw=sunedge, line width=0.8pt] (0,0) circle (\R);
      \node[tny, text=sunedge, anchor=east] at (-0.18,{\Rh+0.20})
           {emission shell, radius $R_\odot+h$};
    
      \draw[ink, line width=0.7pt] (0,-2.65) -- (0,3.60);
      \node[lbl, rotate=90, anchor=south] at (-0.14,-2.00) {plane of sky};
    
      \draw[guide] (0,0) -- (3.65,0);
      \draw[->, ink, line width=0.7pt] (2.90,0) -- (3.63,0);
      \node[tny, anchor=south east, fill=white, inner sep=1pt]
           at (3.65,0.13) {to observer};
      \fill[ink] (0,0) circle (1.3pt) node[lbl, below left=-1pt] {$O$};
      \fill[ink] (\R,0) circle (1.3pt);
      \node[tny, below=2pt, fill=white, inner sep=1pt]
           at (\R,0) {sub-observer point};
    
      \coordinate (P)  at ({\R*cos(\pa)},  {\R*sin(\pa)});
      \coordinate (Pp) at ({\Rh*cos(\pa)}, {\Rh*sin(\pa)});
      \draw[guide] (0,0) -- (Pp);
      \fill[roi] (P) circle (1.8pt);
      \node[tny, text=roi, anchor=north west, fill=white, inner sep=1pt]
           at ($(P)+(0.10,-0.07)$) {region at $(\phi,\theta)$};
    
      \draw[->, ext, line width=1.1pt] (P) -- (Pp);
      \node[lbl, text=ext, anchor=east] at ($(P)!0.55!(Pp)+(-0.05,0.02)$) {$h$};
    
      \draw[ink, line width=0.5pt]
            (0.80,0) arc[start angle=0, end angle=\pa, radius=0.80];
      \node[lbl] at ({1.00*cos(0.5*\pa)},{1.00*sin(0.5*\pa)}) {$\psi$};
      \draw[guide] (P) -- ++(1.10,0);
      \draw[ink, line width=0.5pt]
            ($(P)+(0.47,0)$) arc[start angle=0, end angle=\pa, radius=0.47];
      \node[tny] at ($(P)+({0.64*cos(0.5*\pa)},{0.64*sin(0.5*\pa)})$) {$\psi$};
    
      \draw[guide] (P)  -- (0,{\R*sin(\pa)});
      \draw[guide] (Pp) -- (0,{\Rh*sin(\pa)});
      \fill[roi] (0,{\R*sin(\pa)})  circle (1.5pt);
      \fill[ext] (0,{\Rh*sin(\pa)}) circle (1.5pt);
    
      \draw[ext, decorate, decoration={brace, amplitude=4pt}, line width=0.7pt]
            (-0.33,{\Rh*sin(\pa)}) -- (-0.33,{\R*sin(\pa)});
      \node[tny, text=ext, anchor=east] at (-0.50,{0.5*(\Rh+\R)*sin(\pa)})
           {$\rho = h\sin\psi$};
    
      \node[font=\bfseries\small, anchor=north west] at (-3.75,-3.35)
           {(a) the magnitude of the offset};
      \node[tny, anchor=north west, text width=7.5cm, align=left]at (-3.75,-3.90)
           {A radially outward emission column of height $h$ makes an angle
            $\psi=\arccos(\cos\phi\cos\theta)$ with the line of sight, so only
            its tangential part is visible and the region's apparent position
            shifts by $\rho=h\sin\psi$ --- zero at the disk centre, growing to
            the full $h$ at the limb. };
    \end{scope}
    
    \begin{scope}[xshift=12.2cm]
      \def\Rd{2.45}
      \def\cx{0.883}\def\cy{0.676}
      \def\xa{0.467}\def\xb{1.299}    
      \def\ya{0.356}\def\yb{0.996}    
      \def\dl{0.639}\def\db{0.489}    
    
      \filldraw[fill=sunfill, draw=sunedge, line width=0.8pt] (0,0) circle (\Rd);
      \draw[guide] (-\Rd,0) -- (\Rd,0);
      \draw[guide] (0,-\Rd) -- (0,\Rd);
      \node[tny, anchor=west]  at ({\Rd+0.05},0) {W};
      \node[tny, anchor=south] at (0,{\Rd+0.03}) {N};
    
      \draw[guide] (0,0) -- (2.042,1.565);
      \node[tny, text=gray, anchor=west, rotate=37.4] at (2.07,1.59){radially outward};
      \fill[ink] (0,0) circle (1.3pt);
      \node[tny, below left=-1pt, fill=white, inner sep=1pt] at (0,0) {disk centre};
    
      \fill[ext, opacity=0.16] (\xb,\ya) rectangle ({\xb+\dl},{\yb+\db});
      \fill[ext, opacity=0.16] (\xa,\yb) rectangle ({\xb+\dl},{\yb+\db});
      \draw[ext, densely dashed, line width=0.8pt]
            (\xa,\ya) rectangle ({\xb+\dl},{\yb+\db});
      \draw[roi, line width=1.0pt] (\xa,\ya) rectangle (\xb,\yb);
    
      \draw[ext, line width=0.6pt] (\cx,\cy) -- ({\cx+\dl},\cy)
                                             -- ({\cx+\dl},{\cy+\db});
      \draw[->, ext, line width=1.1pt] (\cx,\cy) -- ({\cx+\dl},{\cy+\db});
      \draw[ink, line width=0.45pt]
            ({\cx+\dl-0.11},\cy) -- ({\cx+\dl-0.11},{\cy+0.11})
                                 -- ({\cx+\dl},{\cy+0.11});
      \fill[roi] (\cx,\cy) circle (1.7pt);
      \node[tny, text=roi, anchor=north east, xshift=1pt, yshift=1pt]
           at (\cx,\cy) {$(\phi,\theta)$};
      \node[tny, text=ext, anchor=north] at ({\cx+0.5*\dl},{\cy-0.03})
           {$\Delta_{\lambda}$};
      \node[tny, text=ext, anchor=west]  at ({\cx+\dl+0.05},{\cy+0.5*\db})
           {$\Delta_{\beta}$};
      \node[tny, text=ext, anchor=south east] at ({\cx+0.55*\dl},{\cy+0.55*\db})
           {$\rho$};
    
      \draw[guide] ({\xb+0.5*\dl},\ya) -- ({\xb+0.5*\dl},-0.58) -- (2.12,-0.58);
      \node[tny, text=ext, anchor=west] at (2.17,-0.58) {$n_{\lambda}$ px};
      \draw[guide] (\xa,{\yb+0.5*\db}) -- (-0.52,{\yb+0.5*\db}) -- (-0.75,1.84);
      \node[tny, text=ext, anchor=east] at (-0.80,1.84) {$n_{\beta}$ px};
    
      \node[font=\bfseries\small, anchor=north west] at (-3.75,-3.35)
           {(b) splitting it between the axes};
      \node[tny, anchor=north west, text width=7.5cm, align=left] at (-3.75,-3.90)
           {$\rho$ points radially outward from the disk centre (it is parallel
            to the region's own apparent position vector) and is the hypotenuse
            of a right triangle whose legs are the per-axis margins,
            $\Delta_{\lambda}=h|\sin\phi\cos\theta|$ and
            $\Delta_{\beta}=h|\sin\theta|$; dividing by the metres-per-pixel
            scale $m$ gives $n_{\lambda}$ and $n_{\beta}$.};
    \end{scope}
    
    \node[anchor=north west, inner sep=7pt, draw=gray!45, line width=0.5pt, rounded corners=2pt, fill=gray!5, text width=16.2cm] at (-0.1,-6.35) {%
      \footnotesize
      In heliocentric Cartesian coordinates with $\hat{z}$ along the line of
      sight, $\hat{x}$ toward solar west and $\hat{y}$ toward solar north, the
      outward radial unit vector at longitude $\phi$ (measured from
      $\textsc{crln\_obs}$) and latitude $\theta$ is
      $\hat{r}=(\sin\phi\cos\theta,\allowbreak\ \sin\theta,\allowbreak\ \cos\phi\cos\theta)$.
      Raising the emission by $h$ along $\hat{r}$ ($h=2.5$~Mm at 1600 and
      1700\,\AA, $h=75$~Mm otherwise) displaces its apparent position by the
      tangential part of $h\hat{r}$, namely
      $\boldsymbol{\Delta}=(h\sin\phi\cos\theta,\allowbreak\ h\sin\theta)$, whence
      \begin{equation*}
        \underbrace{\psi=\arccos\!\big(\cos\phi\cos\theta\big)}
          _{\text{angle from disk centre}}
        \quad
        \underbrace{\rho=\|\boldsymbol{\Delta}\|= h\sin\psi}
          _{\text{apparent offset [m]}}
        \quad
        \underbrace{m=\textsc{cdelt1}\cdot
          \frac{\textsc{rsun\_ref}}{\textsc{rsun\_obs}}}
          _{\text{metres per pixel}}
        \quad
        \underbrace{n_{\lambda}=\Big\lceil
          \frac{h\,|\sin\phi\cos\theta|}{m}\Big\rceil,
          \ \,
          n_{\beta}=\Big\lceil\frac{h\,|\sin\theta|}{m}\Big\rceil}
          _{\text{pixel margin per axis (corrected)}}
      \end{equation*}
      Both $\psi$ and $\rho$ cancel out of the per-axis expressions, so the margins follow from $\phi$, $\theta$ and $h$ alone and $\rho$ is left only as the consistency check $\sqrt{\Delta_{\lambda}^{2}+\Delta_{\beta}^{2}}=h\sin\psi$. Because $\boldsymbol{\Delta}$ is parallel to the region's apparent position vector, the offset is always radial in the image --- a quick check on any implementation. The current code computes $\rho\,|\sin\phi|$ and $\rho\,|\sin\theta|$, the corrected values scaled by $\sin\psi/\cos\theta$ and $\sin\psi$, and so is not radial except at the disk center and on the axes.%
    };
    
    \end{tikzpicture}

    \caption{Computation of the pixel margin $(n_{\lambda},n_{\beta})$
      required to admit a coronal emission height $h$ above the
      photosphere. (a) The radially outward emission column at a region
      lying $\psi$ from the disk centre projects an apparent offset
      $\rho=h\sin\psi$ onto the plane of sky. (b) That offset is split
      between the image axes by its direction cosines, giving
      $\Delta_{\lambda}=h|\sin\phi\cos\theta|$ and
      $\Delta_{\beta}=h|\sin\theta|$, which are converted to pixels
      through the plate scale $m$ and added to the two bounding-box edges
      that face the nearest limb. }
    \label{fig:cutextend}
    
\end{figure}

To create appropriately sized AIA HARP cutouts, we begin by utilizing the minimum bounding rectangle for the bitmap of pixels included in the SHARP parameter calculations. We then enlarge the original HARP bounding boxes in two steps so that coronal loops and emission originating from different atmospheric heights are captured adequately. For each HARP, we first apply a 20\% expansion, capped at 25 arcseconds in both the x and y directions.  In order to ensure that we capture coronal loops, we then further extend the two vertices nearest the solar limb by an amount proportional to their distance from the limb (adapted from \cite{dissauer_properties_2023}). This expansion depends on the wavelength of the input image. For 1600 and 1700\AA, we set the height to 2.5 megameters above the solar surface (following \cite{sanjay_formation_2024}). For all other wavelengths, we use a height of 75 megameters. Figure \ref{fig:cutextend} provides a detailed explanation of this emission-height extension, and Figure \ref{fig:cutout_visual} shows the resulting stack of AIA cutouts for HARP 11149 on May 8th.

\subsection{SDO/AIA Texture Parameter Experiments}\label{section:text_param_exp}
To obtain a reliable set of AIA-based texture parameters suitable for flare forcasting, four features—\emph{Entropy}, \emph{Uniformity}, \emph{Fractal Dimension}, and \emph{Tamura Directionality}— required substantial tuning. Both \emph{Entropy} and \emph{Uniformity} are dependent on the selected bin size used to generate the normalized intensity histogram required in calculations. \emph{Fractal Dimension} depends on the parameters used in the edge detection algorithm, in particular the Gaussian ($\sigma$) chosen for smoothing and the upper threshold applied to select edges based on pixel gradient values. Finally, \emph{Tamura Directionality} is dependent on the number of bins (or quantization) chosen for the histogram of angles and the number of peaks selected within the histogram (please see \cite{Ahmadzadeh2019curated} for more details).

Instead of relying on the tuning parameter settings of \cite{Ahmadzadeh2019curated}, we chose to conduct a similar set of experiments that were used in that work to find optimal settings for our parameter calculations. This was done because of the choice to perform one calculation over the full HARP cutout, as opposed to the 64×64 pixel regions utilized in their work. As was performed in \cite{Ahmadzadeh2019curated}, we defined a broad range of candidate values for each tunable parameter, to effectively explore the parameter space. The initial bounds for each range were taken from those reported by \cite{Ahmadzadeh2019curated}, and were later augmented and refined through experimentation. Tables \ref{tab:fisher_parameter_search} and \ref{tab:bin_search_ranges} describe our final selected range of parameters. 

\begin{table}[t]
        \centering
        \caption{Parameter ranges considered during Fisher-score optimization.
        Entropy and uniformity histogram bins were sampled using
        wavelength-dependent, nonuniform discrete grids.}
        \label{tab:fisher_parameter_search}
        \begin{tabular}{lll}
            \toprule
            Parameter & Search Range & Sampling \\
            \midrule
            Entropy Bins
                & Wavelength Dependent
                & Nonuniform Discrete \\
            Uniformity bins
                & Wavelength Dependent
                & Nonuniform Discrete \\
            Fractal Upper Threshold
                & 0.01 -- 0.16
                & $\Delta = 0.05$ \\
            Fractal Gaussian ($\sigma$)
                & 0.5 -- 4.5
                & $\Delta = 0.5$ \\
            Tamura Quantization
                & 50 -- 95
                & 50, 75, 90, 95 \\
            Tamura Peaks
                & 1 -- 13
                & Odd Integers$^{a}$ \\
            \bottomrule
        \end{tabular}
    
        \vspace{0.5em}
        \begin{minipage}{0.92\linewidth}
            \footnotesize
            $^{a}$The Tamura peak search was extended to 23 for the
            171~\AA\ channel.
        \end{minipage}
    \end{table}

To achieve suitable tuning, we picked five months of observations (covering a broad mix of flaring and non-flaring periods) to perform a small-scale forecasting study. To guide this selection, we relied on the existing SWAN-SF dataset \cite{angryk2020multivariate} labels to calculate the ratio of M/X-class flaring instances relative to $\leq$C-class instances for each month in the original dataset. We then picked the months at the low and high ends of this ratio, along with three months in between that represented intermediate values. The resulting set comprised of July 2010, November 2011, October 2013, February 2014, and October 2014, ordered here from lowest to highest M/X-to-C ratio. 

\begin{table}[t]
    \centering
    \caption{Wavelength-dependent search ranges used for the entropy and
    uniformity histogram bin parameters. Candidate values within each
    interval were sampled nonuniformly.}
    \label{tab:bin_search_ranges}
    \begin{tabular}{cc}
        \toprule
        AIA Wavelength (\AA) & Bin Search Range \\
        \midrule
        94   & 3 -- 155 \\
        131  & 4 -- 490 \\
        171  & 4 -- 4400 \\
        193  & 4 -- 26700 \\
        211  & 4 -- 10300 \\
        304  & 10 -- 2265 \\
        335  & 4 -- 1420 \\
        1600 & 4 -- 2600 \\
        1700 & 4 -- 7080 \\
        \bottomrule
    \end{tabular}
\end{table}
    
Next, we processed the available AIA cutouts (including all nine wavelengths) in SWAN-SF-V2 for the selected months across the full parameter space, evaluating all combinations of feature-specific parameter settings (e.g., \texttt{Tamura Quantization = 1} with \texttt{Tamura Peaks = 1}, \\ \texttt{Tamura Quantization = 1} with \texttt{Tamura Peaks = 3}, etc.). Flare labels were matched to their corresponding HARP regions using the original dataset. Once calculations were complete, and labels assigned, we set up the small-scale study as a binary classification problem. For a particular timestamp and HARP, if a flare $\geq$M-class occurred within the following 24-hours, we would give that timestamp a positive flare class label. All other timestamps were given non-flaring labels. 

\begin{table*}[t]
    \centering
    \caption{Wavelength-dependent parameters selected by maximizing the
    Fisher score. $N_{\mathrm{ent}}$ and $N_{\mathrm{uni}}$ are the
    numbers of histogram bins used for entropy and uniformity,
    respectively. $T_{\mathrm{frac}}$ is the fractal upper threshold,
    $\sigma_{\mathrm{frac}}$ is the Gaussian width used for the fractal
    feature, $Q_{\mathrm{Tam}}$ is the Tamura quantization, and
    $N_{\mathrm{peaks}}$ is the number of Tamura peaks.}
    \label{tab:optimized_feature_parameters}
    \small
    \begin{tabular}{ccccccc}
        \toprule
        Wavelength (\AA)
        & $N_{\mathrm{ent}}$
        & $T_{\mathrm{frac}}$
        & $\sigma_{\mathrm{frac}}$
        & $N_{\mathrm{uni}}$
        & $Q_{\mathrm{Tam}}$
        & $N_{\mathrm{peaks}}$ \\

        \midrule
        94   & 3    & 0.01 & 2.0 & 3  & 75 & 5  \\
        131  & 490  & 0.01 & 1.0 & 4  & 90 & 13 \\
        171  & 4    & 0.01 & 0.5 & 6  & 75 & 23 \\
        193  & 10   & 0.01 & 0.5 & 10 & 90 & 13 \\
        211  & 8    & 0.01 & 0.5 & 12 & 90 & 13 \\
        304  & 2265 & 0.01 & 0.5 & 10 & 95 & 1  \\
        335  & 1420 & 0.01 & 1.0 & 6  & 95 & 1  \\
        1600 & 6    & 0.11 & 1.5 & 10 & 90 & 7  \\
        1700 & 12   & 0.11 & 0.5 & 12 & 95 & 11 \\
        \bottomrule
    \end{tabular}
\end{table*}

Finally, to then analyze the effectiveness of each parameter combination, we calculated their Fisher rank across the dataset. This is a univariate statistic that provides an estimate for the the separation of variances between two distributions (see \cite{bobra2014helioseismic,goodwin_investigating_2024,} for more details). Ideally, it provides a good estimate for a parameters performance when utilized in a full fledged forecasting model (at least in a one dimensional sense). Following this calculation, we then, for each wavelength, select the tuned parameters that provides the highest Fisher rank. The final tuned features are highlighted in Table \ref{tab:optimized_feature_parameters}. These were then utilized to perform the full parameter calculation across the entire SWAN-SF-V2 dataset.

\section{Task-based Dataset Generation} \label{sec:sampling}

As was the case in the original dataset release of \cite{angryk2020multivariate, angryk2020multivariatedata}, our main data product is 7,730 MVTS of solar active region parameters annotated with a collection of co-occurring flares. In this dataset, each MVTS is directly and uniquely associated to a HARP, however many supervised machine learning tasks that this dataset would be useful for do not utilize the full and variable length MVTS. Supervised machine learning tasks can be loosely separated into two categories based on the characteristics of the target variables: classification (if the target variable is discrete) and regression (if the target variable is continuous). For the task of dataset generation, we focus on supervised classification based on discrete flare labels.

An important step towards accelerating machine learning-based solar physics analyses is providing benchmark datasets that are cleaned, partitioned, properly sliced and labeled, as well as consistently balanced based on the number and ratio of flaring (minority) class across partitions. We have already discussed the cleaning procedures applied and will now review the partitioning, slicing, labeling, and balancing procedures. 

\begin{figure} [h!]
   \centering
 	  	\includegraphics[width=0.95\textwidth]{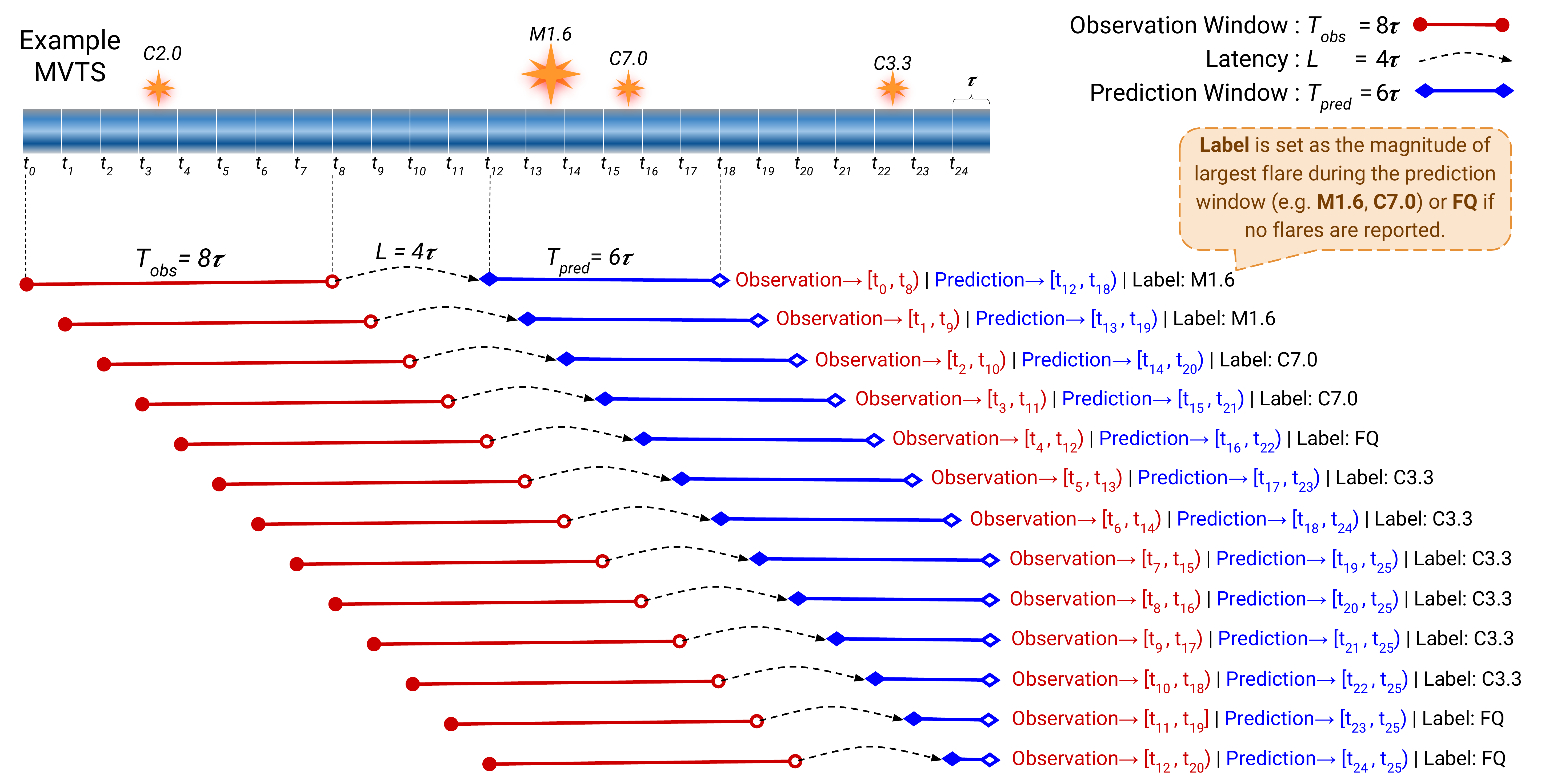} \\
   \caption{Example slicing and labeling of time series, characterized by an elementary time unit of length $\tau$. Time steps ($t_i$) can then be defined at instances corresponding to integer multiples of $\tau$. A slice is a sequence of observation, latency and prediction windows and is labeled as per the largest flare (if any) that occurred during the prediction window. }
   \label{fig:sampling}
 \end{figure}

\subsection{Partitioning:} 
The first step in creating a machine learning model is to determine the task, and therefore, to specify the target classes. Target classes are determined using flare intensity threshold criteria. For a common binary classification schema, where M- or X-class flares ($\geq$M1.0) are considered flaring and lower magnitude flares ($<$M1.0) and flare quiet instances are considered non-flaring, target class specification will use a single threshold value [M1.0]. For creating a 4-class classification schema (e.g., B-class or lower ($\leq$B9.9), C-class ($\geq$C1.0 and $\leq$C9.9), M-class ($\geq$M1.0 and $\leq$M9.9), and X-class ($\geq$X1.0)), we can use [C1.0, M1.0, X1.0] as the threshold criteria. Different threshold criteria can be produced for different tasks.

It is important to remember that large flares (M- or X-class), which have the greatest impacts on the space environment and are thus the most commonly targeted in predictive analyses, are scarce. In our dataset, we have 730 M-class flares and only 50 X-class flares, corresponding to a mere $\sim$6.8\% of all flare records included in the dataset. Among 7,730 MVTS, only 72 contain X-class flares and 452 have M-class flares, corresponding to a slim $\sim$5.9\% of the total. 

In machine learning applications the creation of validation datasets is usually performed by holding out parts of datasets one or more times, so that the models can learn from the training sets and generalize on samples they have never seen before. Given the scarcity of major flaring examples, the previous release of the dataset in \cite{angryk2020multivariate, angryk2020multivariatedata} proposed a more robust validation strategy for machine learning applications in solar flare prediction: \emph{time-segmented stratification}. In the previous efforts, the stratification method separated the dataset into unequal time intervals (partitions). These different intervals, included similar total numbers of major flares (i.e., members of the minority class) in each partition. For example, in a partitioned MVTS with balanced minority class populations, a total of 780 M- and X-class flares split between five partitions would give rise to rough totals of 156 M-/X-class flares per partition. However, it was determined that this method lead to partitions with a varied imbalance ratio between M- and X-class flare labeled samples and those labeled as not in the separate partitions. This lead to inconsistent results depending upon which partition was chosen for training and which was chosen for testing/validation.   

So, in this release we have opted to continue with a time-segmented stratification as it prevents correlations between different time series segments stemming from the same MVTS being in both the training and testing partitions. However, the time-segmentation will instead be based on which quarter of the year the HARP was observed.  With this method, we can still ensure (1) that we have non-overlapping time segments in each partition, so that the training and testing samples rely on different MVTS, (2) still have a relatively equal number of minority instances across all partitions as their frequency of occurrence tends to follow the solar cycle and not earths orbit around the sun.

\subsection{Slicing and Labeling:} 
The next step after partitioning the dataset is to methodically slice and label MVTS based on a desired prediction scenario. To achieve that, we utilize the same methodologies described for \cite{angryk2020multivariate, angryk2020multivariatedata}, where the concept of \emph{observation window}, \emph{latency}, and \emph{prediction window} were introduced. The observation window length ($T_{obs}$) is used to determine the duration of time series slices for the sampling of predictive parameters. To label each of these slices with the appropriate flare occurrence, latency ($L$) and prediction window ($T_{pred}$) lengths are specified. Here latency represents the time interval from the issuing of a forecast (end of the observation window) to its coming into effect at the start of the prediction window. The prediction window is then the interval of validity of that forecast.

For a time series slice (i.e., observation, latency, and prediction windows) starting at $t_i$, the observation window corresponds to the interval at $[t_i,t_i\!\!+\!\!T_{obs})$. The prediction window corresponds to the interval at $[t_i\!+\!T_{obs}\!+\!\!L, t_i\!+\!T_{obs}\!+\!L\!+\!T_{pred})$. Each instance (slice) are then labeled with the magnitude of the largest flare (if any) that occurred in that HARP during the prediction window. A schematic, exemplary scenario of slicing and labeling for a MVTS is presented in Fig.~\ref{fig:sampling}.

\chapter{Data Schema}
\begin{table}[t]
\caption{Summary and categorization of the parameters included in our dataset.}
\label{tab:records}

\centering
\scriptsize

\renewcommand{\arraystretch}{1.1}
\setlength{\tabcolsep}{6pt}

\newcommand{\colblock}[1]{%
    \begin{tabular}[t]{@{}>{\ttfamily}c@{}}
        #1
    \end{tabular}%
}

\begin{adjustbox}{max width=\linewidth,center}

\begin{tabular}{@{} c c c c @{}}

\toprule


\textbf{HARP Meta Data}
&
\textbf{Magnetic Field Parameters (Table~\ref{tab:mfp})}
&
\textbf{AIA Texture Parameters}
&
\textbf{Flare Meta Data}
\\

\cmidrule(r){1-1}
\cmidrule(lr){2-2}
\cmidrule(lr){3-3}
\cmidrule(l){4-4}


\colblock{
    HARP\_NUMBER \\
    OBS\_START \\
    LL\_LON \\
    LL\_LAT \\
    UR\_LON \\
    UR\_LAT \\
}

&

\begin{tabular}[t]{@{}
>{\centering\arraybackslash}p{1.55cm}
>{\centering\arraybackslash}p{1.55cm}
>{\centering\arraybackslash}p{1.55cm}
>{\centering\arraybackslash}p{1.55cm}
@{}}

\textit{Parameter}
&
\textit{Uncertainty}
&
\textit{Parameter}
&
\textit{Uncertainty}
\\

\cmidrule(lr){1-2}
\cmidrule(lr){3-4}

\ttfamily TOTFZ   & \ttfamily ERRTOTFZ
& \ttfamily MEANJZD & \ttfamily ERRJZ \\

\ttfamily TOTFX   & \ttfamily ERRTOTFX
& \ttfamily TOTUSJZ & \ttfamily ERRUSI \\

\ttfamily TOTFY   & \ttfamily ERRTOTFY
& \ttfamily MEANALP & \ttfamily ERRALP \\

\ttfamily TOTBSQ  & \ttfamily ERRTOTBSQ
& \ttfamily MEANJZH & \ttfamily ERRMIH \\

\ttfamily EPSZ    & \ttfamily ERREPSZ
& \ttfamily TOTUSJH & \ttfamily ERRTUI \\

\ttfamily EPSX    & \ttfamily ERREPSX
& \ttfamily ABSNJZH & \ttfamily ERRTAI \\

\ttfamily EPSY    & \ttfamily ERREPSY
& \ttfamily SAVNCPP & \ttfamily ERRJHT \\

\ttfamily USFLUX  & \ttfamily ERRVF
& \ttfamily MEANPOT & \ttfamily ERRMPOT \\

\ttfamily MEANGAM & \ttfamily ERRGAM
& \ttfamily TOTPOT & \ttfamily ERRTPOT \\

\ttfamily MEANGBT & \ttfamily ERRBT
& \ttfamily MEANSHR & \ttfamily ERRMSHA \\

\ttfamily MEANGBZ & \ttfamily ERRBZ
& \ttfamily SHRGT45 & -- \\

\ttfamily MEANGBH & \ttfamily ERRBH
& \ttfamily R\_VALUE & -- \\

\end{tabular}

&

\begin{tabular}[t]{@{}c@{}}

\colblock{
    ENTROPY \\
    MEAN \\
    STD \\
    FRACTAL\_DIM \\
    SKEWNESS \\
    KURTOSIS \\
    UNIFORMITY \\
    REL\_SMOOTHNESS \\
    T\_CONTRAST \\
    T\_DIRECTION
}

\end{tabular}

&

\colblock{
    START\_TIME \\
    PEAK\_TIME \\
    END\_TIME \\
    GOES\_CLASS \\
    CONFIDENCE \\
    HAS\_GOES \\
    HAS\_SSW \\
    HAS\_XRT \\
    HARP\_VOTE \\
    MAX\_PAIR\_DISTANCE\_DEG
}

\\
\midrule
\multicolumn{4}{@{}l@{}}{%
\textbf{Slicing Meta Data}: 
\texttt{SLICE\_ID}, 
\texttt{SLICE\_START\_TIME}, 
\texttt{OBS\_END\_TIME}, 
\texttt{PRED\_START\_TIME}, 
\texttt{PRED\_END\_TIME}, 
\texttt{PEAK\_CLASS}
}
\\
\bottomrule

\end{tabular}

\end{adjustbox}

\end{table}
As described throughout this paper, SWAN-SF-V2 is a validated collection of shear-weighted and unweighted magnetic field parameters, their uncertainties, and AIA texture parameters for HARPs observed from May 2010 through December 2025. For each HARP and associated timestamp, we obtain 6 parameters related to the \textbf{HARP Meta Data}, 46 parameters related to the \textbf{Magnetic Field}, 90 related to \textbf{AIA}, and 10 related to the \textbf{Flare Metadata} (if applicable).  Table \ref{tab:records} summarizes the parameters in each of these four categories.

Within the \textbf{HARP Meta Data} category we have information about the observed \texttt{HARP\_NUMBER}, timestamp (\texttt{OBS\_START}), and HARP bounding box (\texttt{LL\_LON}, \texttt{LL\_LAT}, \texttt{UR\_LON}, \texttt{UR\_LAT}).

Within the \textbf{Magnetic Field} category there are 24 parameters, with corresponding shear-weighted versions of them denoted by a prefix \texttt{W\_}. There are also 22 associated parameter uncertainties, with corresponding shear-weighted versions. Within the \textbf{AIA} category there are 10 texture parameters that are calculated independently across the 94, 131, 171, 193, 211, 304, 335, 1600, and 1700~\AA{} passbands, making 90 in total. Wavelengths are appended to each AIA parameter name (e.g., \texttt{MEAN\_94}).

Finaly, in the \textbf{Flare Meta Data} category, we include details about the flare linked to the HARP, specifically its \texttt{START\_TIME}, \texttt{PEAK\_TIME}, \texttt{END\_TIME}, and \texttt{GOES\_CLASS}. We also provide the \texttt{CONFIDENCE} label for the event (see Section \ref{sec:flare_verification} for additional information), as well as the catalogs used for verification (\texttt{HAS\_GOES}, \texttt{HAS\_SSW}, \texttt{HAS\_XRT}). In addition, this category contains the \texttt{HARP\_VOTE} (e.g., 2-1 indicates two matches within the same HARP and one match in a different HARP) and the largest angular separation between catalog positions, given in degrees (\texttt{MAX\_PAIR\_DISTANCE\_DEG}).

The MVTS files are stored in Parquet file format, where all of the files in one directory represent one DataFrame that the appropriate Parquet reader would be pointed at. There are 4 directories for each type of MVTS we have constructed a dataset for. In this release, as was done for the original dataset, we have files that have the full length timeseries of an AR listed as a series of observations, grouped by \texttt{HARP\_NUMBER}, and ordered by timestamp (\texttt{OBS\_START}) within each group.  We also have sliced and labeled data that is meant as a machine learning ready dataset, with the observation window set to 60 time steps (bounded by \texttt{SLICE\_START\_TIME}, \texttt{OBS\_END\_TIME}), prediction window set as 120 time steps (bounded by \texttt{PRED\_START\_TIME}, \texttt{PRED\_END\_TIME}), and steps between samples set as 5 timesteps. In this version of the dataset there is an additional column of \texttt{SLICE\_ID} to designate how repeated entries differ, and \texttt{PEAK\_CLASS} to indicate what the strongest GOES flare class that occurs over the prediction window for that slice. 

The two different versions of dataset partitions are currently available at \url{https://dmlab.cs.gsu.edu/solar/swan-sf/}. These will be transferred to Harvard Dataverse, with an accompanying DOI, once a finalized version is completed after peer review.  



\section*{Acknowledgements}
\noindent This work was supported by NASA's Space Weather Science Application Research-to-Operations-to-Research program grant \#80NSSC22K0272, Multidomain Reusable Artificial Intelligence Tools program grant \#80NSSC23K1026, and Heliophysics Artificial Intelligence/Machine Learning-ready Data program grant \#80NSSC24K0238. Author Goodwin's contributions were also supported by NASA FINESST grant \#80NSSC23K1639.


\section*{Competing interests}
The authors declare no competing financial interests

\clearpage
\fancyfoot[C]{References}
\fancypagestyle{plain}{%
\fancyfoot[C]{References}
}
\printbibliography

\end{document}